\documentclass[10pt,a4paper]{article}
\usepackage{amsfonts}
\usepackage{times}
\usepackage{amsbsy}
\usepackage{amssymb}
\usepackage{amsmath}
\usepackage[load=named]{siunitx}
\usepackage{graphicx}
\usepackage{vmargin}
\usepackage{setspace}
\usepackage[latin1]{inputenc}
\usepackage[numbers]{natbib}
\usepackage{color}

\setmargrb{20mm}{20mm}{13mm}{20mm}

\begin{document}

\title{External Tonehole Interactions in Woodwind Instruments}
\author{Antoine Lefebvre and Gary P. Scavone \\
%EndAName
Computational Acoustic Modeling Laboratory, \\
Centre for Interdisciplinary Research in Music Media and Technology (CIRMMT)%
\\
Schulich School of Music, McGill University,\\
555 Sherbrooke Street West, Montréal, Québec H3A 1E3, Canada\\
\vspace{1cm} \\
Jean Kergomard \\
LMA, CNRS, UPR 7051, Aix-Marseille Univ, Centrale Marseille, \\
F-13402 Marseille Cedex 20, France}
\maketitle

\begin{abstract}
The classical Transfer-Matrix Method (TMM) is often used to calculate
the input impedance of woodwind instruments. However, the TMM ignores
the possible influence of the radiated sound from toneholes on other
open holes. In this paper a method is proposed to account for external
tonehole interactions. We describe the Transfer-Matrix Method with
external Interaction (TMMI) and then compare results using this
approach with the Finite Element Method (FEM) and TMM, as well as with
experimental data.  It is found that the external tonehole
interactions increase the amount of radiated energy, reduce slightly
the lower resonance frequencies, and modify significantly the response
near and above the tonehole lattice cutoff frequency. In an appendix,
a simple perturbation of the TMM to account for external interactions
is investigated, though it is found to be inadequate at low frequencies
and for holes spaced far apart.

%The results of simulations with the FEM, as well as experimental measurements, are presented and compared to the calculations with the method presented in this paper, confirming that the external tonehole interactions play a significant role in woodwind instruments.

\end{abstract}

\doublespacing

\section{Introduction}

\label{intro}

A method to accurately and efficiently estimate the input impedance and
resonance frequencies of woodwind instruments is of primary importance. The
Transfer-Matrix Method (TMM) is typically used for this purpose (see e.g. %
\citet{plitnik_numerical_1979,causse_input_1984,keefe_woodwind_1990}),
because of its simplicity and efficiency, and it is the basis of software
used by some instrument makers, such as RESONANS or BIAS. This method
ignores internal interactions due to the coupling between the evanescent
modes of nearby discontinuities as well as external interactions, which
exists because the radiation impedance of each open tonehole is influenced
by the radiation of sound from other toneholes. The problem of the response
of woodwind instruments with external tonehole interactions was stated in a
complete form by \citet{leppington_theory_1982} and a method of solution was
proposed by \citet{kergomard_1989} using the mutual radiation impedance
proposed by \citet{pritchard_mutual_1960}. This method is based on the TMM
for internal propagation with modified open tonehole radiation impedances to
account for interactions (referred to as TMMI). Preliminary experimental
results were obtained by \citet{helmut} for the case of holes spaced far
apart in a pipe. No other validation of the method has been proposed since
reference \cite{kergomard_1989}.

In this paper, we investigate the effect of external tonehole
interactions in woodwind instruments with the TMMI. The first goal is
to determine the validity of the method by comparing the results of
TMMI calculations with Finite Element Method (FEM) simulations and wth
measurements. We also compare these results with TMM calculations to
show that some of the discrepancies are explained by the external
interactions.

The second goal is to apply the proposed TMMI method to the case of
woodwind instruments in order to determine the importance of the
effect on their acoustical properties and to judge whether or not it
is necessary to account for those interactions when calculating the
input impedance of woodwind instruments for design purposes.

The theory of the TMM and TMMI, as well as the details of the FEM, are
reviewed in the next section, and the presentation of the TMMI is completed
in Appendix A for a general model of open holes. This is followed by the
presentation of the results for a tube with a regular array of holes (Sec.\ %
\ref{results}), then results for a saxophone and a clarinet (Sec.\ \ref{SC})
and finally the conclusions (Sec.\ \ref{conclusion}). In Appendix B the
possibility to use a perturbation approach for the TMM is investigated.

\section{Background}

\subsection{The TMM}

The transfer matrix method (TMM) provides an efficient means for calculating
the input impedance of a hypothetical air column %
\citep{plitnik_numerical_1979, causse_input_1984, keefe_woodwind_1990}. With
the TMM, a geometrical structure is approximated by a sequence of
one-dimensional segments, such as cylinders, cones, and closed or open
toneholes, and each segment is represented by a transfer matrix (TM) that
relates its input to output frequency-domain quantities of pressure ($P$)
and volume velocity ($U$). The multiplication of these matrices yields a
single matrix which must then be multiplied by an appropriate radiation
impedance, $Z_{rad}$, at its output as:
\begin{equation}
\begin{bmatrix}
P_{in} \\
U_{in}%
\end{bmatrix}%
=\left( \prod_{i=1}^{n}\mathbf{T}_{i}\right)
\begin{bmatrix}
Z_{rad}U_{out} \\
U_{out}%
\end{bmatrix}%
.
\end{equation}%
The input impedance is then calculated as $Z_{in}=P_{in}/U_{in}$,
without need to know $U_{out}.$

The theoretical expression of the transfer matrix of a cylinder is:
\begin{equation}
\mathbf{T}_{cyl}=%
\begin{bmatrix}
\cosh {(\Gamma L)} & Z_{0}\sinh {(\Gamma L)} \\
Z_{0}^{-1}\sinh {(\Gamma L)} & \cosh {(\Gamma L)}%
\end{bmatrix}%
,  \label{M2}
\end{equation}%
where $Z_{0}=\rho c/\pi a^{2}$, $\Gamma =j\omega /c+(1+j)\alpha $ is the
complex propagation constant, $\rho $ is the air density, $c$ the speed of
sound, and $\omega $ the angular frequency. Losses are represented by $%
\alpha $, which depends on the radius $a$ of the cylindrical pipe and varies
with the square root of the lossless wavenumber $k=\omega /c$:
\begin{equation}
\alpha =(CST/a)\sqrt{k},
\end{equation}%
where $CST$ is a constant that depends of the properties of air:
\begin{equation}
CST=\sqrt{\ell _{v}/2}(1+(\gamma -1)/\sqrt{\Pr }),
\end{equation}%
$\ell _{v}=\mu /\rho c$ is the characteristic length of viscous
effects, $\mu$ is the dynamic viscosity, $\Pr $ the Prandtl number and
$\gamma$ the ratio of specific heats. The formulas above for $Z_{0}$
and $\Gamma$ are sufficiently accurate for the present study.

The transfer matrix of a conical waveguide is (see %
\citet{chaigne_acoustique_2008}):

\begin{equation}
\mathbf{T}_{cone}=%
\begin{bmatrix}
(a_{2}/a_{1})\cos {(k_{c}L)}-\sin {(k_{c}L)}/{kx_{1}} & jZ_{c}\sin {(k_{c}L)}
\\
Z_{c}^{-1}\left[ j(1+(k^{2}x_{1}x_{2})^{-1})\sin {(k}_{c}{L)}%
+(x_{1}^{-1}-x_{2}^{-1})\cos {(k}_{c}{L)}/jk\right] & (a_{1}/a_{2})\cos {(k}%
_{c}{L)}+\sin {(k}_{c}{L)}/{kx_{2}}%
\end{bmatrix}%
,  \label{eq:transmission_matrix_cone}
\end{equation}%
where $a_{1}$ and $a_{2}$ are the radii at the input and output planes,
respectively, and $x_{1}$ and $x_{2}$ are the distances between the apex of
the cone and the input and output planes, $Z_{c}=\rho c/(\pi a_{1}a_{2})$
and $k_{c}=-j\Gamma $ is the complex wavenumber. In this case, losses are
evaluated at the equivalent radius \citep{chaigne_acoustique_2008}:
\begin{equation}
a_{eq}=L\frac{a_{1}}{x_{1}}\frac{1}{\ln {(1+L/x_{1})}}.
\end{equation}

The transfer matrix of a tonehole is defined as:
\begin{equation}
\mathbf{T}_{hole}=\left(
\begin{array}{cc}
1 & Z_{a}/2 \\
0 & 1%
\end{array}%
\right) \left(
\begin{array}{cc}
1 & 0 \\
Z_{s}^{-1} & 1%
\end{array}%
\right) \left(
\begin{array}{cc}
1 & Z_{a}/2 \\
0 & 1%
\end{array}%
\right)  \label{eq:tm_tonehole_shunt_series}
\end{equation}%
where $Z_{a}$ is the series impedance and $Z_{s}$ the shunt impedance. These
impedances have different values in the open and closed state. The
calculation of these impedances has been the subject of many articles (%
\citet{nederveen_corrections_1998, dubos_1999, dalmont_experimental_2002,
lefebvre_2012}) and the reader is referred to those papers for the
appropriate formulas.

\subsection{The TMMI}

\subsubsection{Structure of the computation}

The radiation impedance of each tonehole on a woodwind instrument is
influenced by the sound radiated from other holes. A method of solution to
account for such interactions was proposed by \citet{kergomard_1989}.
%The method is described with more details here.
It can be used for any bore shape by making use of the classical TMM with
modifications for the matrices located between open toneholes. It gives
identical results to the TMM if interactions are neglected (by specifying
null mutual radiation impedances). That is, the geometry is discretized
identically, with both closed tonehole and open tonehole series impedance
terms $Z_{a}$ represented as in the TMM.
% The method is described with more details in this section.
% In this section, we present the details of this method.

We assume an instrument with $N$ openings (embouchure hole, toneholes, open
end), where the indices of the openings range from $n=1$ to $N$. The
pressure $P_{n}^{rad}$ at opening $n$ is related to the acoustic flow $%
U_{n}^{rad}$ radiating out of hole $n$ by the following matrix relationship:

\begin{equation}
\mathbf{P}^{rad}=\mathbb{Z}\mathbf{U}^{rad},  \label{eq:pu}
\end{equation}%
where we define the vector $\mathbf{P}^{rad}$ of the pressures $P_{n}^{rad}$
and the vector $\mathbf{U}^{rad}$ of the flow rates $U_{n}^{rad}.$ $\mathbb{Z%
}$ is the radiation impedance matrix, which includes the effect of external
interactions. The precise values of the different elements are difficult to
determine. The self radiation impedances are the diagonal elements. The
validity of this expression comes directly from the integral form of the
Helmholtz equation if the Green function is chosen to satisfy the Neumann
boundary conditions on the tube (see e.g. Eq. (7.1.17) in %
\citet{morse_ingard_1968}, see also \citet{leppington_theory_1982}). As a
consequence, the equations used by Keefe \citep{keefe_acoustic_1983} are
erroneous (see Eqs. A1a to A2b). The content of this matrix is explained in
Sec.\ \ref{sec:impedance_matrix}.

A complete description of the planar mode propagation inside the tonehole
chimney is possible, as explained in Appendix A. If the height is smaller
than the wavelength, this can be simplified as:%
\begin{eqnarray}
P_{n} &=&P_{n}^{rad}+B_{n}U_{n}, \\
U_{n}^{rad} &=&U_{n},  \notag
\end{eqnarray}%
where $P_{n}$ is the pressure at the hole inside the air column and
$B_{n}$ is the impedance of the total acoustic mass of the hole (see
Appendix A). This approximation is possible for the frequency range of
the present study. Using the diagonal matrix $\mathbb{B}$, we write:%
\begin{equation}
\mathbf{P=P}^{rad}+\mathbb{B}\mathbf{U}=(\mathbb{Z+B)}\mathbf{U}  \label{K1}
\end{equation}

\begin{figure}[t]
\begin{center}
\includegraphics{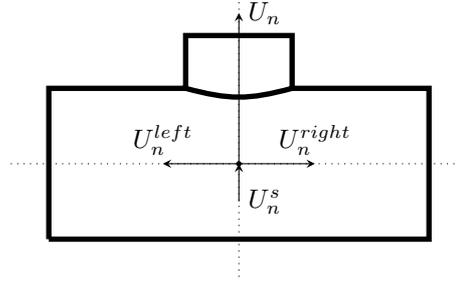}
\end{center}
\caption{Diagram of the flow contributions in Eq.~\eqref{eq:flows}.}
\label{fig:flows}
\end{figure}

An alternative equation relating the pressures and flows due to propagation
inside the instrument can be derived for each hole $n$. As illustrated in
Fig.~\ref{fig:flows}, the sum of the flow $U_{n}$ radiating out of the
tonehole, the flow $U_{n}^{right}$ entering the tonehole section on the
right, and the flow $U_{n}^{left}$ entering the tonehole section on the left
is equal to the flow source $U_{n}^{s}$ (which is discussed hereafter). This
flow conservation equation can be written as:
\begin{equation}
U_{n}^{s}=U_{n}+U_{n}^{left}+U_{n}^{right},  \label{eq:flows}
\end{equation}%
where we note that the flow on the left is defined in the reverse direction.

This equation can be written in a matrix form as:
\begin{equation}
\mathbf{U}^{s}=\mathbf{U}+\mathbf{U}^{left}+\mathbf{U}^{right},
\label{eq:us}
\end{equation}%
where $\mathbf{U}^{s}$ is the flow source vector. The sum of the left and
right internal flows for each section is related to the pressures as
\begin{equation}
\mathbf{U}^{left}+\mathbf{U}^{right}=\mathbb{Y}\mathbf{P},
\label{eq:uleftright}
\end{equation}%
where $\mathbb{Y}$ is the admittance matrix, which is described in Sec.~\ref%
{sec:admittance_matrix}.

By combining Eqs.~\eqref{K1}, \eqref{eq:us} and \eqref{eq:uleftright} , we
obtain the solution:
\begin{equation}
\mathbf{U}=\left[ \mathbb{I}+\mathbb{Y(}\mathbb{Z+B)}\right] ^{-1}\mathbf{U}%
^{s},  \label{k12}
\end{equation}%
where $\mathbb{I}$ is the identity matrix.

\subsubsection{Source Vector of Flow Rates $\mathbf{U}^{s}$}

For such a calculation, the flow-source vector $\mathbf{U}^{s}$ needs to be
known. Generally speaking, the reader can imagine a small loudspeaker
located inside the pipe at the abscissa of each hole, providing a flow rate $%
U_{n}^{s}$. Clearly, the resonator of a musical instrument is passive and
such sources do not exist. All transfer functions between two acoustic
quantities at every point in the space of the passive system are fully
determined and the unique problem is the choice of a reference. A solution
is to use as a reference the flow rate on the left at the first open tonehole of
the instrument (from the part of the instrument that does not have any open
holes), $-U_{1}^{left}$. This quantity can be regarded as a source, if a
source is defined as a fixed (or forced) quantity (in the sense of the
Thevenin theorem). In the absence of an active source, Eq.~\eqref{eq:us} for
this hole becomes:.%
\begin{equation*}
-U_{1}^{left}=U_{1}+U_{1}^{right}.
\end{equation*}%
Thus, we can replace Eq.~\eqref{eq:us} for the first open hole by the following:%
\begin{equation}
U_{1}^{s}=U_{1}+U_{1}^{right}(=-U_{1}^{left})  \label{k221}
\end{equation}%
Otherwise $U_{n}^{s}=0$ for $n\neq 1$. In this way, we can compute all
quantities with respect to $U_{1}^{s}$, i.e. the ratio of all quantities to $%
U_{1}^{s}.$ The input impedance can be easily deduced from the knowledge of $%
U_{1}^{s}$ and $P_{1}$, the ratio $U_{1}^{s}/P_{1}$ being the input
admittance $Y^{up}$ of the part of the system with open toneholes. Therefore
the input impedance is classically computed by projecting the impedance $%
1/Y^{up}$ at the input of the instrument, or by using a transfer matrix
relationship. Then all quantities can be calculated with respect to the
input flow rate $U_{0}^{right}$ if necessary, where the index 0 refers to
quantities at the input plane of the system. For reed instruments, this
quantity is related to the input pressure by a time-domain nonlinear
characteristic.\footnote{%
For flute-like instruments, it should be possible to choose the flow rate $%
U_{1}$ exiting from the mouthpiece, which is the first open hole, as a
source. However a complete nonlinear model needs to consider a
pressure-difference source (i.e. a force source) near the edge, and it is
necessary to add an equation in order to compute the flow rates radiating
from the holes with respect to this source.}

\subsubsection{Radiation Impedance Matrix $\mathbb{Z}$}

\label{sec:impedance_matrix}

The self-radiation impedances $Z_{nn}$ are approximately known (see e.g. %
\citet{dalmont_radiation_2001}). On the other hand, at low frequencies, the
mutual radiation impedance $Z_{nm}$ (when $n\not=m$) is, assuming that holes
radiate as monopoles (see \citet[Eq.~(17)]{pritchard_mutual_1960}):
\begin{equation}
Z_{nm}=jk\rho c\frac{e^{-jkd_{nm}}}{2\pi d_{nm}},  \label{k15}
\end{equation}%
where $d_{nm}$ is the distance between toneholes $n$ and $m$. More
closely spaced toneholes have a larger mutual radiation impedance. As
the mutual impedance is complex, both reactive and dissipative effects
are expected. The factor 2 in the denominator corresponds to the
radiation of a monopole into a half space. At very low frequencies a
factor 4 would be more logical because the radiation is into a
complete space, but empirically we noted that at higher frequencies,
when the effect of interaction is especially important, a factor 2 is
more suitable.  It is difficult to determine the best approximation
for the radiation impedance (see e.g. \citet{dalmont_radiation_2001,
  dalmont_experimental_2002}).

Therefore, the impedance matrix $\mathbb{Z}$ is a full matrix. The mutual
impedance may be neglected by using a diagonal matrix $\mathbb{D}$ with self
impedance only, in which case the results are identical to those of the TMM.

\subsubsection{Admittance Matrix $\mathbb{Y}$}

\label{sec:admittance_matrix}

The propagation of planar sound waves between two toneholes 1 and 2 can be
described by classical transfer matrices:
\begin{equation}
\begin{bmatrix}
P_{n} \\
U_{n}^{right}%
\end{bmatrix}%
=%
\begin{bmatrix}
A_{n} & B_{n} \\
C_{n} & D_{n}%
\end{bmatrix}%
\begin{bmatrix}
P_{n+1} \\
-U_{n+1}^{left}%
\end{bmatrix}%
,  \label{M3}
\end{equation}%
where the transfer matrix is the multiplication of the transfer matrices of
each segment located between the two open toneholes, including any closed
toneholes. As explained above, the series impedances $Z_{a}$ of the open
toneholes can be accounted for by including them in the transfer matrix,
one-half on each side.

This matrix can be written in the form of an admittance matrix:
\begin{equation}
\begin{bmatrix}
U_{n}^{right} \\
U_{n+1}^{left}%
\end{bmatrix}%
=%
\begin{bmatrix}
Y_{n} & Y_{\mu ,n} \\
Y_{\mu ,n} & Y_{n}^{\prime }%
\end{bmatrix}%
\begin{bmatrix}
P_{n} \\
P_{n+1}%
\end{bmatrix}%
,  \label{eq:admittance_matrix}
\end{equation}%
The parameters of this matrix are related to those of the transfer matrix: $%
Y_{n}=D_{n}/B_{n}$, $Y_{n}^{\prime }=A_{n}/B_{n}$ and $Y_{\mu ,n}=-1/B_{n}$,
which assumes that $A_{n}D_{n}-B_{n}C_{n}=1$, the condition for reciprocity.
The right and left flows at one tonehole section $n$ become:
\begin{equation}
U_{n}^{right}=Y_{n}P_{n}+Y_{\mu ,n}P_{n+1}  \label{eq:flowright}
\end{equation}%
and
\begin{equation}
U_{n}^{left}=Y_{\mu ,n-1}P_{n-1}+Y_{n-1}^{\prime }P_{n}  \label{eq:flowleft}
\end{equation}%
Thus, Eq.~\eqref{eq:flows} can be expanded to:
\begin{equation}
U_{n}^{s}=U_{n}+Y_{\mu ,n-1}P_{n-1}+(Y_{n-1}^{\prime }+Y_{n})P_{n}+Y_{\mu
,n}P_{n+1}.  \label{eq:flowsum}
\end{equation}%
The coefficients of this equation define the admittance matrix $\mathbb{Y}$,
which is tridiagonal. The first and last equations have to be modified
because there is either no previous opening or no next opening. The last
opening is located at the far end of the instrument, so that $%
U_{N}^{right}=0 $ and Eq.~\eqref{eq:flowsum} becomes simply:
\begin{equation}
U_{N}^{s}=U_{N}+Y_{\mu ,N-1}P_{N-1}+Y_{N-1}^{\prime }P_{N},
\end{equation}%
where $U_{N}$ is the flow rate radiated at the end of the tube.

For the first opening, we use Eq.~\eqref{k221}, where we can set $U_{1}^{s}$%
, the first entry of the flow source vector, to any value. Then, using Eq.~%
\eqref{k12}, solving the problem gives the flow vector $\mathbf{U}$. The
pressure vector $\mathbf{P}$ can be deduced with Eq.~\eqref{eq:pu}.

\subsection{Finite Element Calculations}

The evaluation of the input impedance of woodwind instruments using the FEM
involves constructing a 3D model of the air column surrounded by a radiation
sphere and the solution of the Helmholtz equation for a number of selected
frequencies. The body of the instrument itself is considered to be rigid.
The mesh occupies the volume inside and outside the instrument. Curved
third-order Lagrange elements are used.

The input impedance (or reflectance) is evaluated from the FEM solution by
evaluating the relationship of pressure and volume flow (or traveling-wave
components of pressure) at the input plane of the system (see also %
\citet{lefebvre_2012}). The surrounding spherical radiation domain uses a
second-order non-reflecting spherical boundary condition on its surface, as
described by \citet{bayliss_boundary_1982}. Further discussion on this topic
can be found in \citet{tsynkov_numerical_1998} and %
\citet{givoli_high_order_2003}.

Thermoviscous boundary layer losses may be approximated with a special
boundary condition such as presented by \citet{cremer1948acoustic} and, more
recently, \citet{bossart_hybrid_2003} or \citet{kampinga_performance_2010}.
The boundary condition can be written as a specific acoustic admittance:
\begin{equation}
Y_{wall}=-\frac{v_{n}}{p}=\frac{1}{\rho c}\sqrt{jk\ell _{v}}\left[ \sin ^{2}{%
\theta }+(\gamma -1)/\sqrt{\Pr }\right] ,  \label{eq:admittance_losses}
\end{equation}%
where $v_{n}$ is the normal velocity on the boundary and $\theta $ is the
angle of incidence of the plane wave. The angle of incidence may be
calculated from $\cos {\theta }=\hat{n}\cdot \hat{v}/||\hat{v}||$, where the
normal vector $\hat{n}$ is of unit length. This is solved iteratively. The
lossless problem is solved first, then the admittance on the boundary is
calculated from the normal velocity of the solution and the problem is
solved again. This is repeated until convergence is found.

The properties of air at \SI{25}{°C} are used for all the simulation cases.
See \citet{causse_input_1984} for the equations used to calculate those
values.

The reflectance $R(f)=p_{-}/p_{+}$ (ratio of the reflected to incident
pressure) is obtained from the simulation results. A cylindrical segment is
added before the input plane of the object under study. The pressures $p_{a}$
and $p_{b}$ at two points on the centerline of this cylindrical segment, at
distances $a$ and $b$ from the input plane, are extracted and the
reflectance is calculated as:
\begin{equation}
R=\frac{e^{-\Gamma b}-H_{ba}e^{-\Gamma a}}{H_{ba}e^{\Gamma a}-e^{\Gamma b}},
\end{equation}%
where $H_{ba}=p_{b}/p_{a}$ is the transfer function between the two
pressures and $\Gamma $ is as previously defined. A singularity in this
equation exists when the distance is half of the wavelength. The reduced
impedance can then be calculated with $\overline{Z} = (1+R)/(1-R) $.

This method to calculate the reflectance was inspired by the two-microphones
transfer function method of impedance measurement. It is worth mentioning
that the impedance could also be calculated as $\overline{Z_{in}}%
=p_{in}/\rho cv_{in}$, where the pressure and velocity are extracted
directly at the input plane. When validating this approach using a
cylindrical pipe, it was found that the results did not match theory as well
as with the two-point method \cite{lefebvre_PhDthesis}.

\begin{figure}[t]
\begin{center}
\includegraphics[width=246pt]{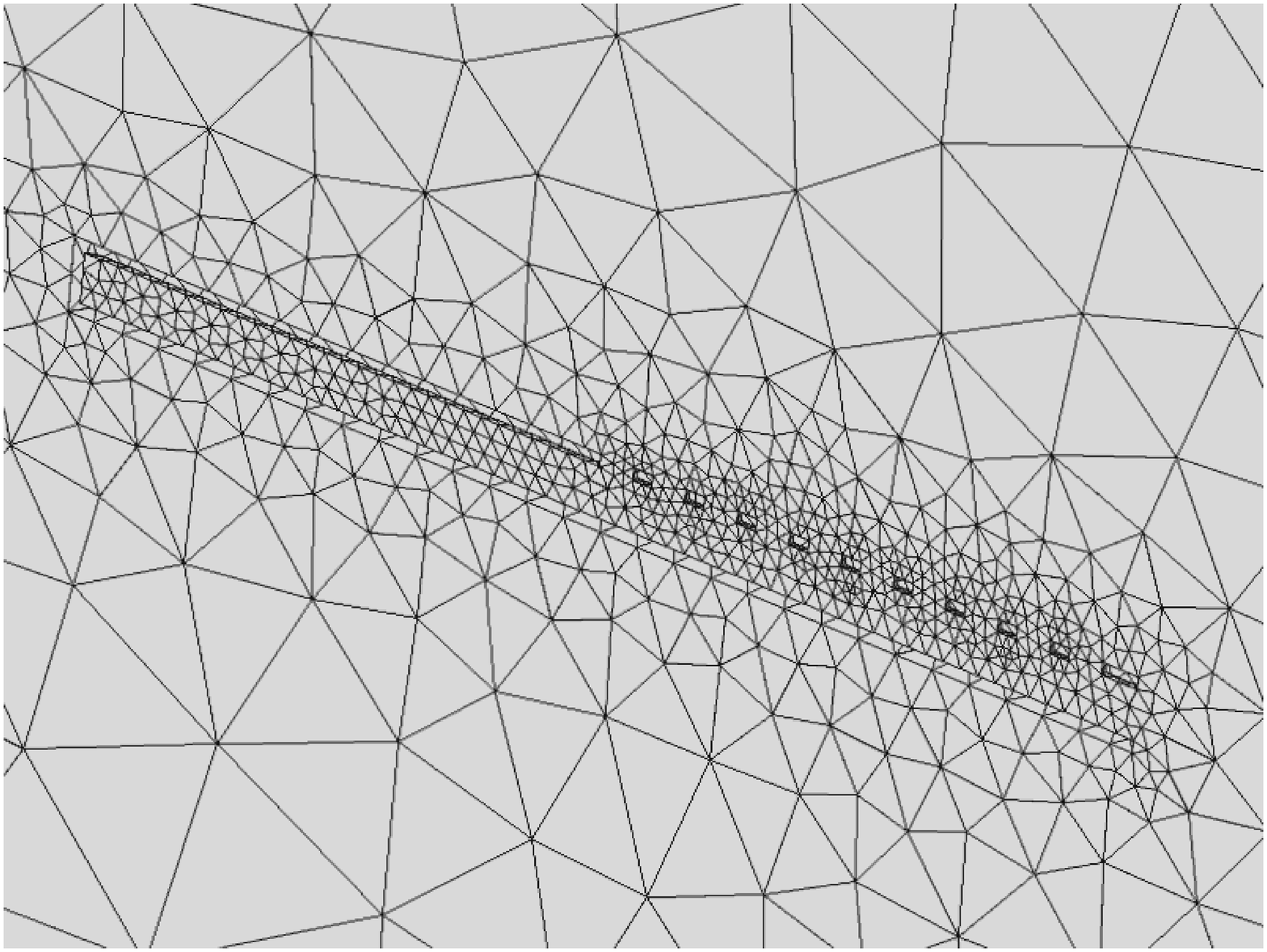}
\end{center}
\caption{The mesh of a pipe with 10 open toneholes.}
\label{fig:mesh}
\end{figure}

\section{Results for a Pipe with a Regular Array of Holes\label{results}}

\begin{figure}[t]
\begin{center}
\includegraphics[width=246pt]{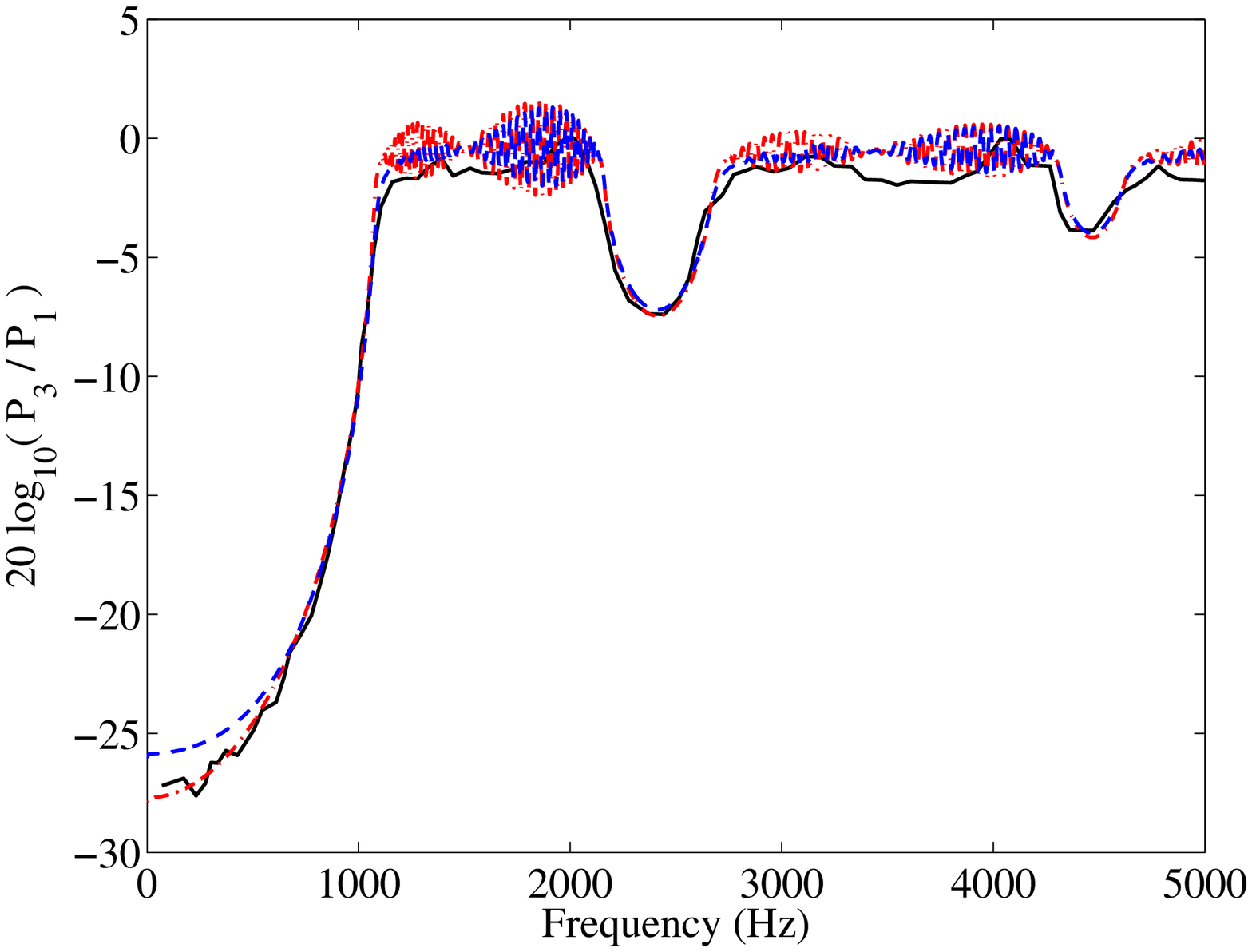} %
\includegraphics[width=246pt]{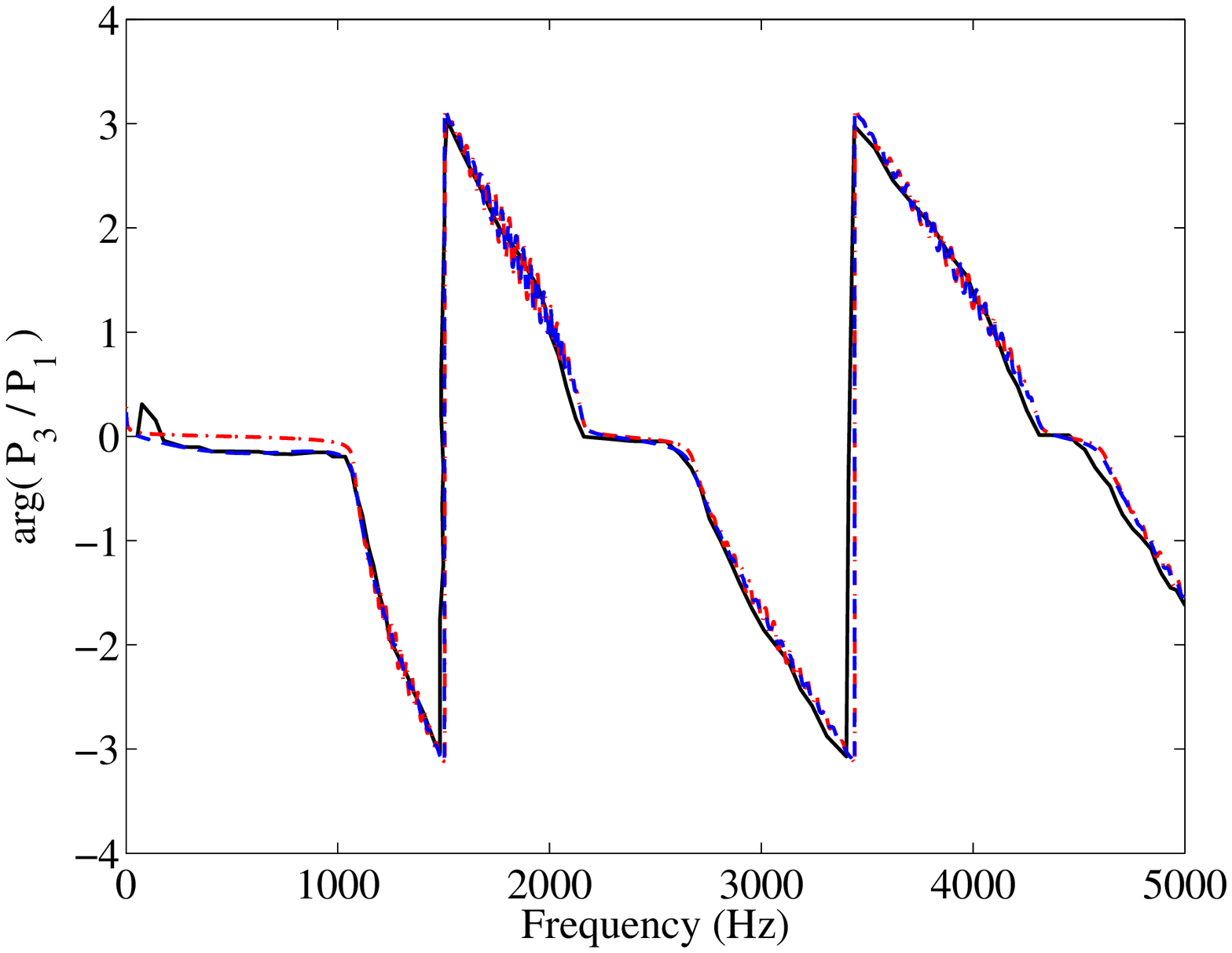}
\end{center}
\caption{Modulus and argument of the transfer function between the internal
pressure at hole 3 and 1: experimental results (black solid), theoretical
calculation without external interactions (TMM, red dash-dotted) and with
interactions (TMMI, blue dashed).}
\label{fig:p3_p1}
\end{figure}

\begin{figure}[t]
\begin{center}
\includegraphics[width=246pt]{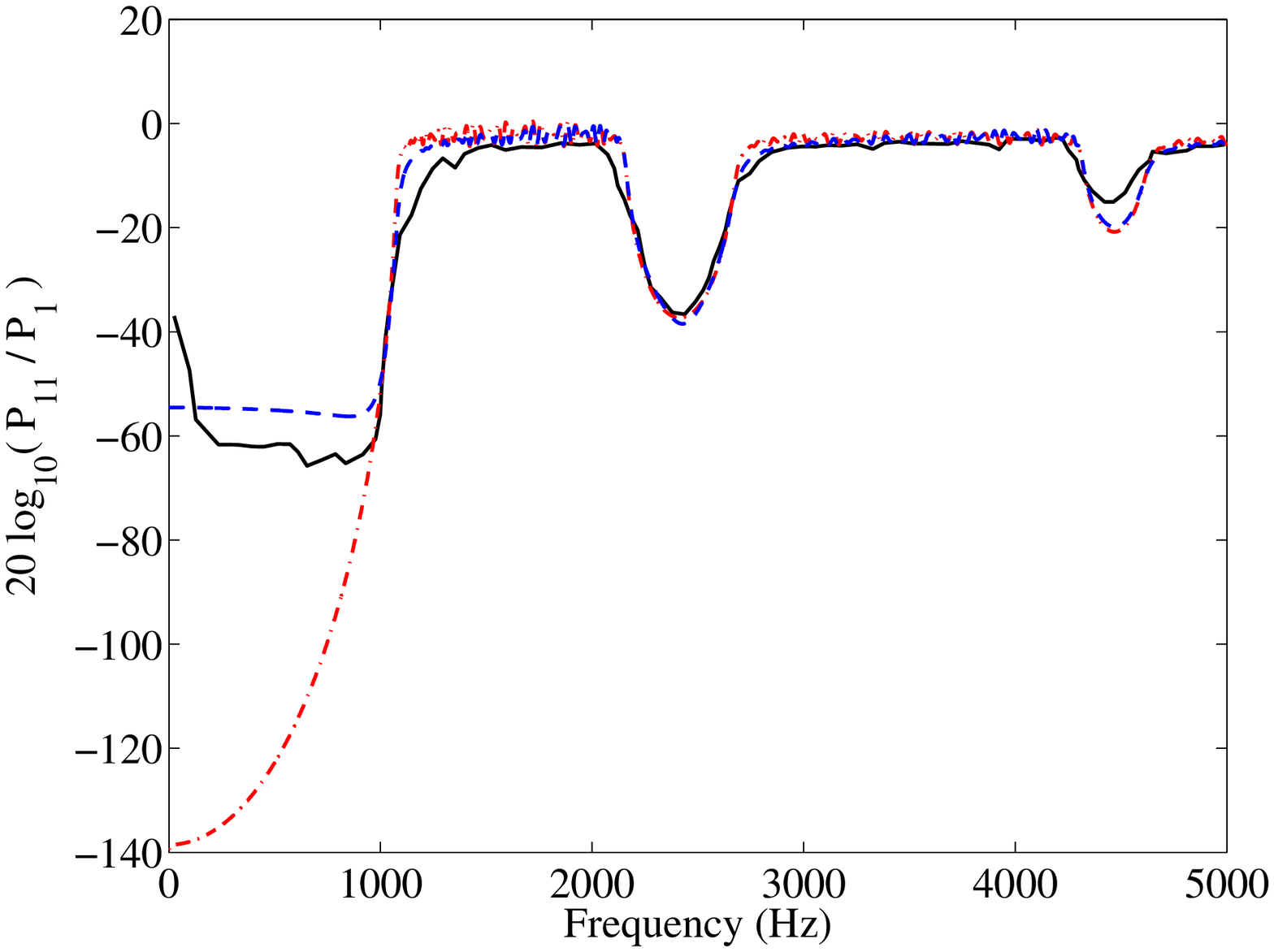} %
\includegraphics[width=246pt]{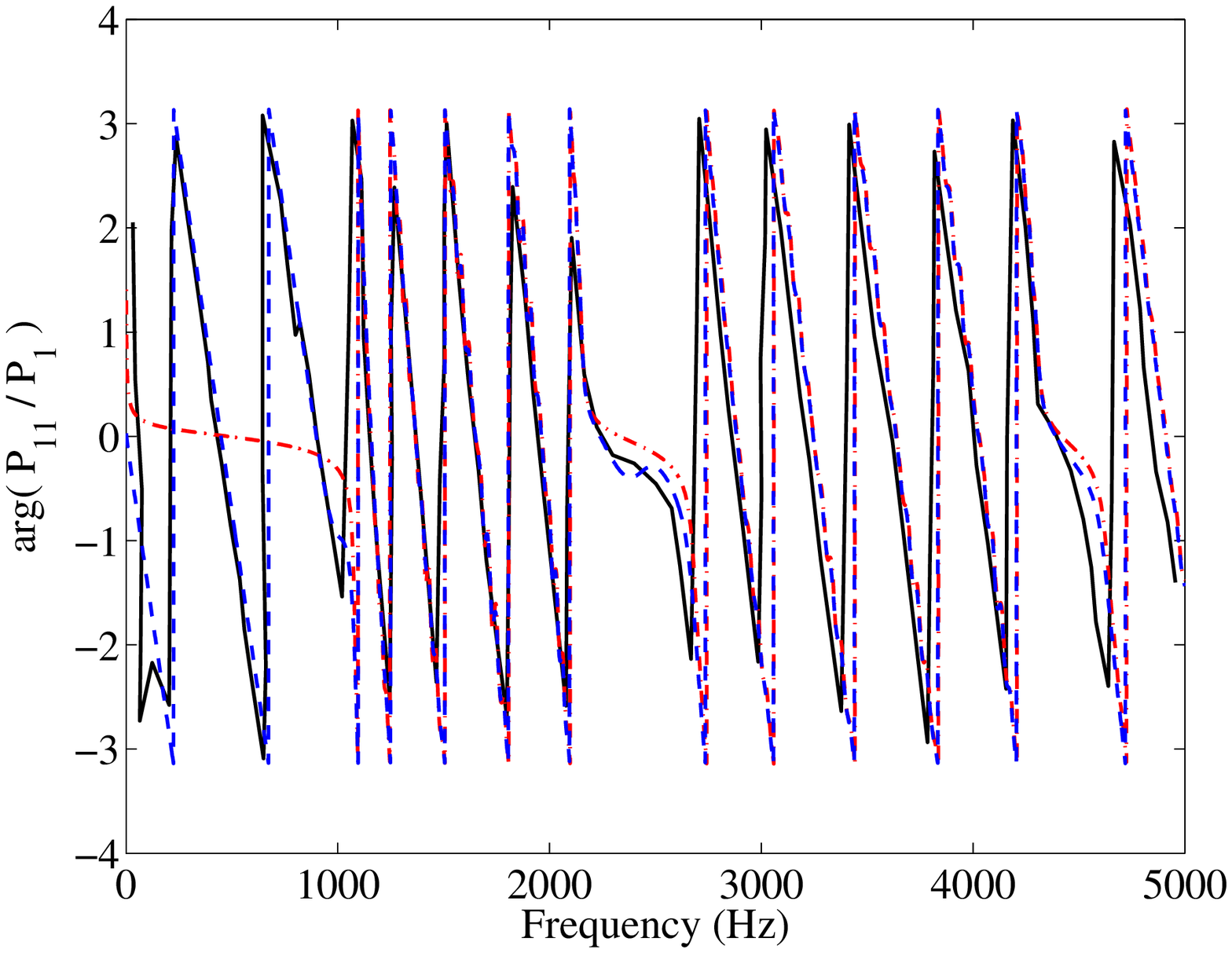}
\end{center}
\caption{Modulus and argument of the transfer function between the internal
pressure at hole 11 and 1: experimental results (black solid), theoretical
calculation without external interaction (TMM, red dash-dotted) and with interaction
(TMMI, blue dashed).}
\label{fig:p11_p1}
\end{figure}

External tonehole interactions were studied experimentally by %
\citet{helmut} during an internship in Le Mans, France. The experiment
involved measuring the internal pressure at the position of the holes
on a tube with an array of widely spaced toneholes. The distances
between the toneholes was much larger than what is found on woodwind
instruments but the conclusion remains applicable to some extent. A
cylindrical pipe of 4 meters length with an internal diameter of
15.3\thinspace mm was drilled with 47 holes of 8.7\thinspace mm
diameter regularly separated by 8\thinspace cm.  The far end of the
pipe was rigidly capped. The wall thickness was 3\thinspace mm, and
the temperature 20$°$C. The excitation (white noise signal) was
provided by a loudspeaker at the input of the tube. The internal
pressure was measured at the positions of hole 1, 3 and 11 using
1/4-in B\&K microphones mounted flush with the pipe wall opposite the
toneholes. The transfer functions with respect to the pressure at the
first tonehole were calculated using an HP analyser and a computer. The
results are shown in Figs.~\ref{fig:p3_p1} and \ref{fig:p11_p1} in
comparison to theoretical calculations with and without interactions.

For frequencies lower than the cutoff frequency of a tonehole lattice,
the sound is exponentially attenuated inside the waveguide when
interactions are ignored, whereas the external pressure is inversely
proportional to distance. Therefore, the acoustic pressure coming from
outside of the toneholes located farther down an instrument becomes
stronger than the pressure coming from inside the instrument. In
Fig. \ref{fig:p3_p1}, it appears that the effect of the external
interactions is negligible for the 3rd tonehole because the pressure
coming from inside remains important, but in Fig. \ref{fig:p11_p1} the
internal pressure has sufficiently decayed at the 11th hole such that
the external sound field dominates. The phase curve is very
instructive: when interactions are ignored the phase shift is very
small, indicating evanescent waves, while when interactions are taken
into account, the phase variation is linear, indicating (spherical)
traveling waves.

Therefore the effect of interaction is extremely strong for this case
(widely spaced holes and low frequencies). This is the reason why a
perturbation method starting from the TMM cannot be used: this idea is
investigated in Appendix B. Unfortunately, the convergence is limited
to pass bands (above the first cutoff), and this method, used by
\citet[p.115]{nederveen_corrections_1998}, cannot be used for stop
bands.

For frequencies higher than the cutoff frequency of the tonehole lattice,
the internal pressure is no longer exponentially attenuated and the effect
of interactions is limited to a smoothing of the response.

\begin{figure}[t]
\begin{center}
\includegraphics[width=246pt]{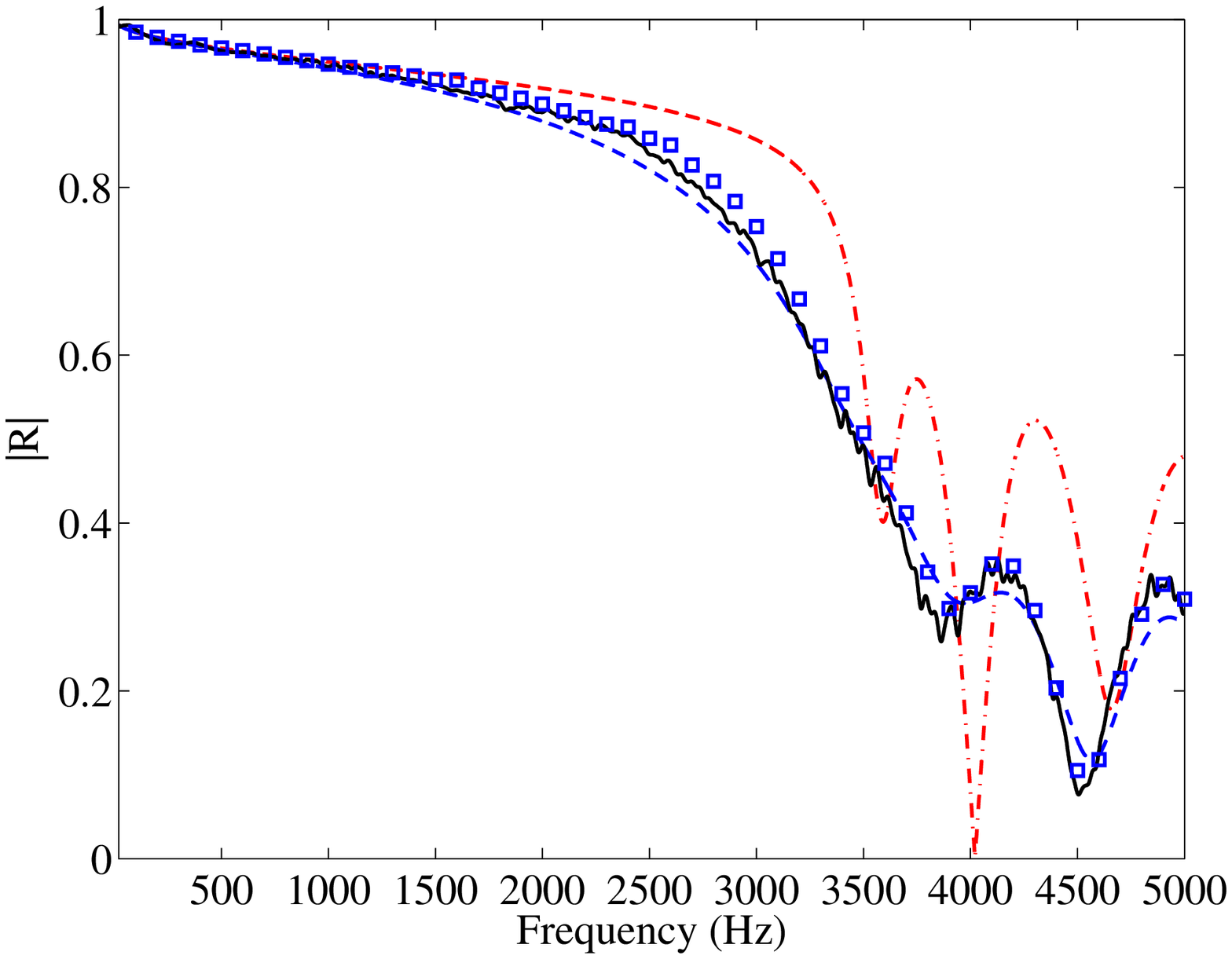} %
\includegraphics[width=246pt]{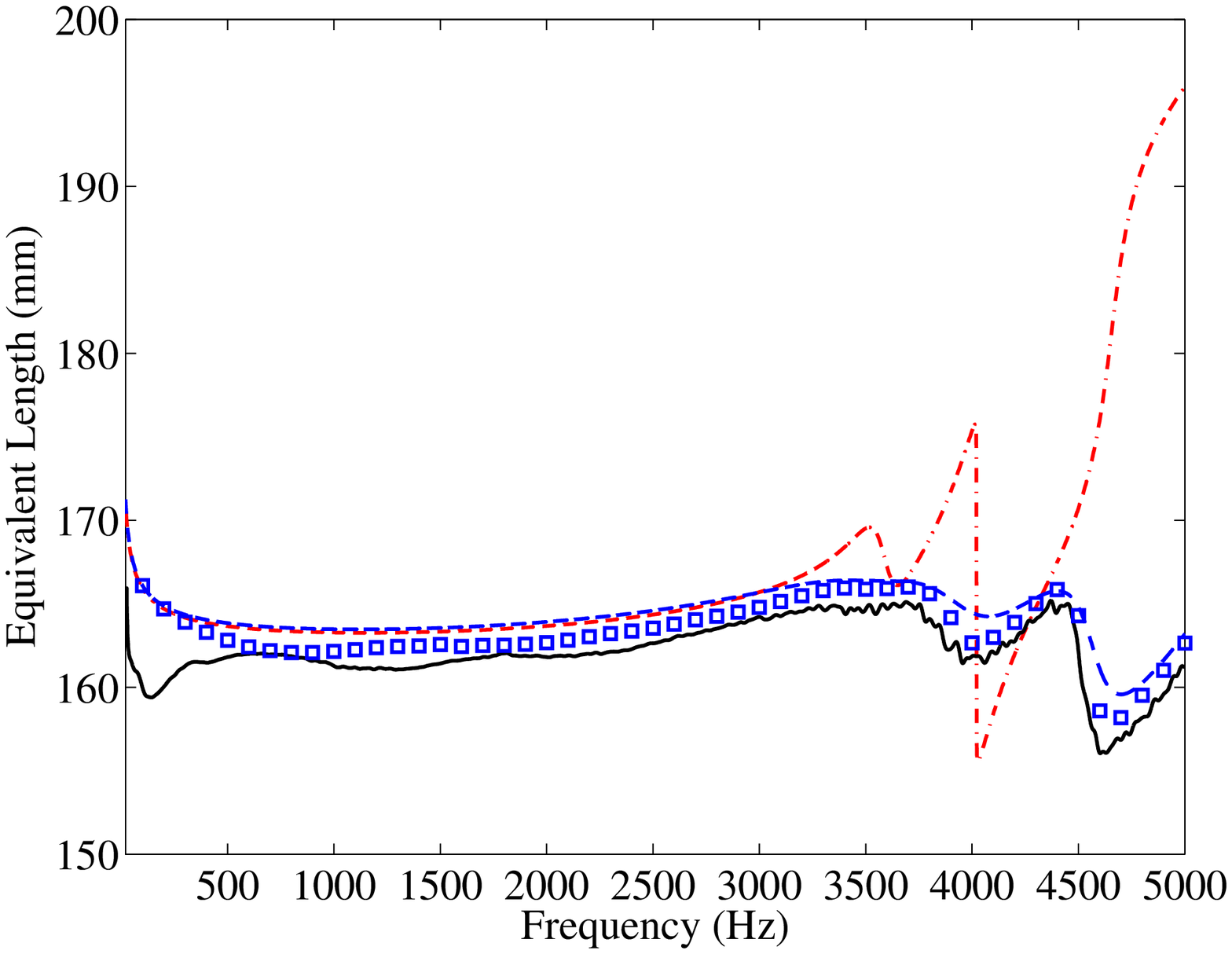}
\end{center}
\caption{Magnitude of the reflectance (left) and equivalent length (right)
of a pipe with 10 toneholes: experimental results (black solid), FEM simulation
results (squares), calculations with external interactions (TMMI, blue dashed)
and calculations without interactions (TMM, red dash-dotted).}
\label{fig:proto1_R}
\end{figure}

In order to better understand the impact of the existence of external
interactions on the playing characteristics of woodwind instruments,
the influence of the external interactions on the input impedance (or
reflectance) of a pipe with an array of closely spaced toneholes was
studied. The pipe was 303\thinspace mm in length, with a
12.7\thinspace mm diameter and 3.2\thinspace mm wall thickness.  It
was drilled with 10 holes of 9.5\thinspace mm diameter equally spaced
by 15\thinspace mm, starting at a position of 153\thinspace mm from
the input end. The reflectance of that pipe was obtained with the
proposed calculation method and compared to simulation results with
the FEM, to experimental measurements (see description in the next
section) and to calculations with the classical TMM. The magnitude of
the reflectance and the equivalent length are plotted in
Fig.~\ref{fig:proto1_R}.

An important observation is that the FEM results closely match the
experimental measurement. % for the magnitude of the reflectance.
This significantly increases our confidence in both the FEM and the
experiment. For the equivalent length, the measurement error appears
to be larger, particularly for the lower frequencies (below
1000\thinspace Hz).  There also seems to be a slight systematic error
of a few millimeters. The proposed TMMI calculation to account for
external interactions clearly gives better results than the TMM. The
deep minima in the magnitude of the reflectance and the large increase
in equivalent length in the higher frequencies completely disappear
when interactions are included. The overall shape of the curves
resemble the measured and simulated ones, even though some
discrepancies remain. In the lower frequencies, the magnitude of the
reflectance is reduced by the external interactions, which indicates a
higher radiation efficiency. In the higher frequencies, the minima in
the magnitude of the reflectance is not as low as in the measurement
and is not located exactly at the same frequency. Small discrepancies
also exist in the equivalent length, though they appear to be on the
order of the measurement errors. In the lower frequencies, the
external interactions increase the equivalent length slightly compared
to predictions of the TMM. The toneholes on the pipe are located very
close to each other, so that the evanescent modes excited near each
discontinuity interact with those of adjacent toneholes, that is, the
propagation of sound between toneholes is not planar, as assumed in
the proposed method. This phenomena is one likely cause of the
remaining discrepancies. Another is that the model of the mutual
interaction assumes that each tonehole is a monopole. In spite of
those simplifications, the proposed method gives improved
results. Most of the discrepancies between the classical TMM and the
measurements are explained by the presence of external tonehole
interactions.

Generally speaking, Fig.~\ref{fig:proto1_R} exhibits a major feature of pass
bands: external interaction yields a significant reduction of oscillations
with frequency, i.e. a reduction of the standing wave amplitude. This
feature was stated by \citet{kergomard_1989}. In Appendix B, a theoretical
justification is given, allowing the following interpretation:

\begin{itemize}
\item without interaction, there is reflection at the end, with standing
waves inside the lattice;

\item without interaction, standing waves imply the existence of extrema of
flow rate, the different holes radiating at different levels;

\item the holes radiating strongly have an important influence on the holes
radiating weakly, thus there is a kind of equalization of radiation by the
different holes, thus a diminution of the apparent standing wave ratio (SWR);

\item finally at the input of the lattice there is a diminution of the
reflection coefficient.
\end{itemize}

A consequence is the reduced height of the impedance peaks above the cutoff
frequency and a reduction in the radiation directivity lobes in the backward
direction \citep{kergomard_1989}. An analysis of the clarinet cutoff
frequencies taking into account the external interaction can be found in %
\citep{moers}.

\section{Results for a Saxophone and a Clarinet\label{SC}}

A precise computational FEM model of a complete music instrument is
difficult to create and requires significant computation time to
solve.  Thus, the TMMI model can provide a faster and easier numerical
technique which provides satisfactory results for real instruments
with complicated geometry, despite the fact that the theoretical
description of the toneholes with key pads is overly simplified. In
general, we can at least expect that qualitative effects are well
represented. Results of the TMM, TMMI, and measurements for an alto
saxophone and a clarinet are presented in this section.  The saxophone
and clarinet were measured with their mouthpieces removed.

\subsection{Input impedance measurements}

The input impedance measurements were made with a multi-microphone
system, as described in \citep{lefebvre:fa2011}.  A JBL 2426 horn
driver is attached to one end of the probe and six PCB Piezotronics
condenser microphones (model 377B10) with preamplifiers (model 426B03)
are mounted flush with the inner probe wall at 30\thinspace mm,
60\thinspace mm, 100\thinspace mm, 150\thinspace mm, 210\thinspace mm,
and 330\thinspace mm from the input plane of the pipe.  The
microphones are connected to a PCB Piezotronics signal conditioner
(model 483C30) and then to a computer through an RME Fireface 800
audio interface.  The signals are sampled at 48 kHz.  The system is
excited with a repeated logarithmically-swept sine tone of length
32768 samples and the resulting responses to this signal are averaged
in the time-domain, with the first response being discarded.  The
spectral analysis is performed with an FFT size of 32768, giving a
frequency resolution of 1.46 Hz.  The pressure spectra at each
microphone are used to solve for the forward and backward traveling
waves in the system \citep{jang:microphone}, effectively measuring the
reflectance of an attached object.  The probe is calibrated with three
non-resonant loads, as described in \citep{dickens:calibration},
though a time-windowing technique \citep{kemp:waveseparation} is used
for the quasi-infinite length pipe.

A cylindrical pipe of 60 cm was measured with this system and compared
to measurements using a CTTM impedance probe \cite{roux:capteurz}, as
well as the TMM.  All results were within 2 dB and 1\% frequency
accuracy at impedance magnitude extrema between 50--2000 Hz (the
frequency range of interest in the following sections).

\subsection{Saxophone}

\begin{figure}[tb]
\begin{center}
\includegraphics[width=246pt]{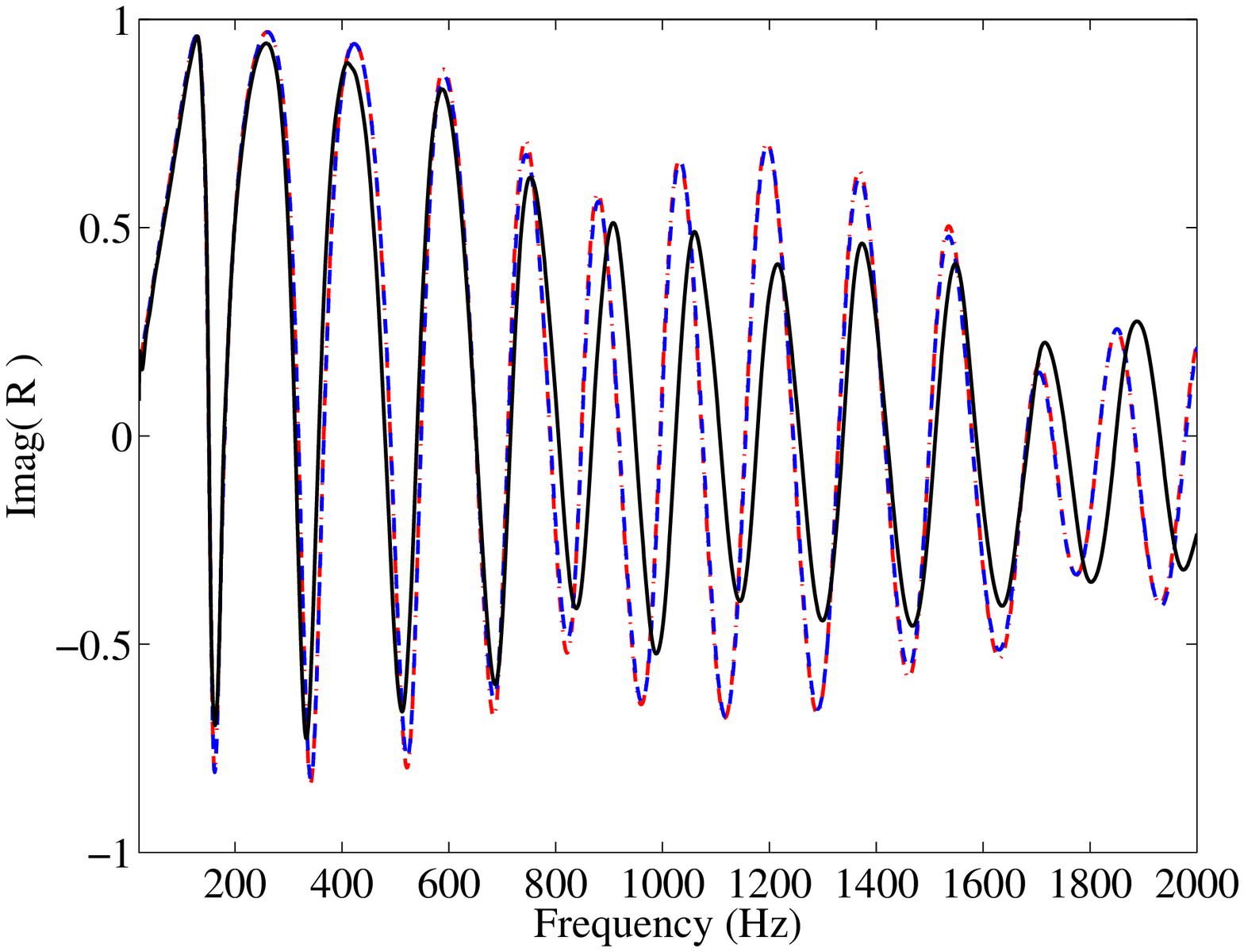} %
\includegraphics[width=246pt]{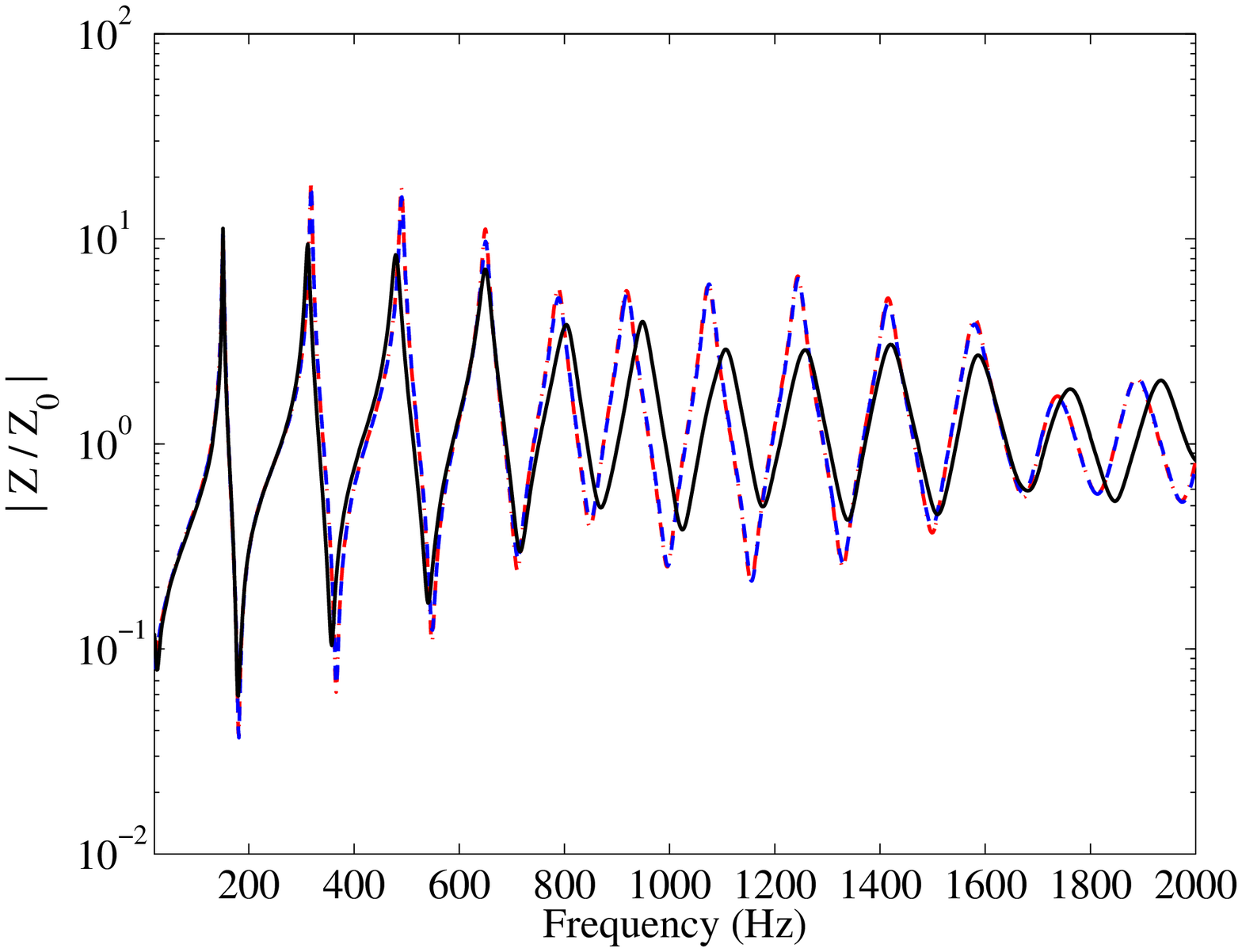}
\end{center}
\caption{Imaginary part of the reflectance (left) and magnitude of the
  impedance (right) of an alto saxophone with a low $B$ fingering:
  experimental results (black solid) calculations with external
  interactions (blue dashed) and calculations with the TMM (red
  dash-dotted). Because of there are few open holes, the interaction
  effects are very small and the two curves TMM and TMMI are barely
  distinguishable.}
\label{fig:saxo_B}
\end{figure}

\begin{figure}[t]
\begin{center}
\includegraphics[width=246pt]{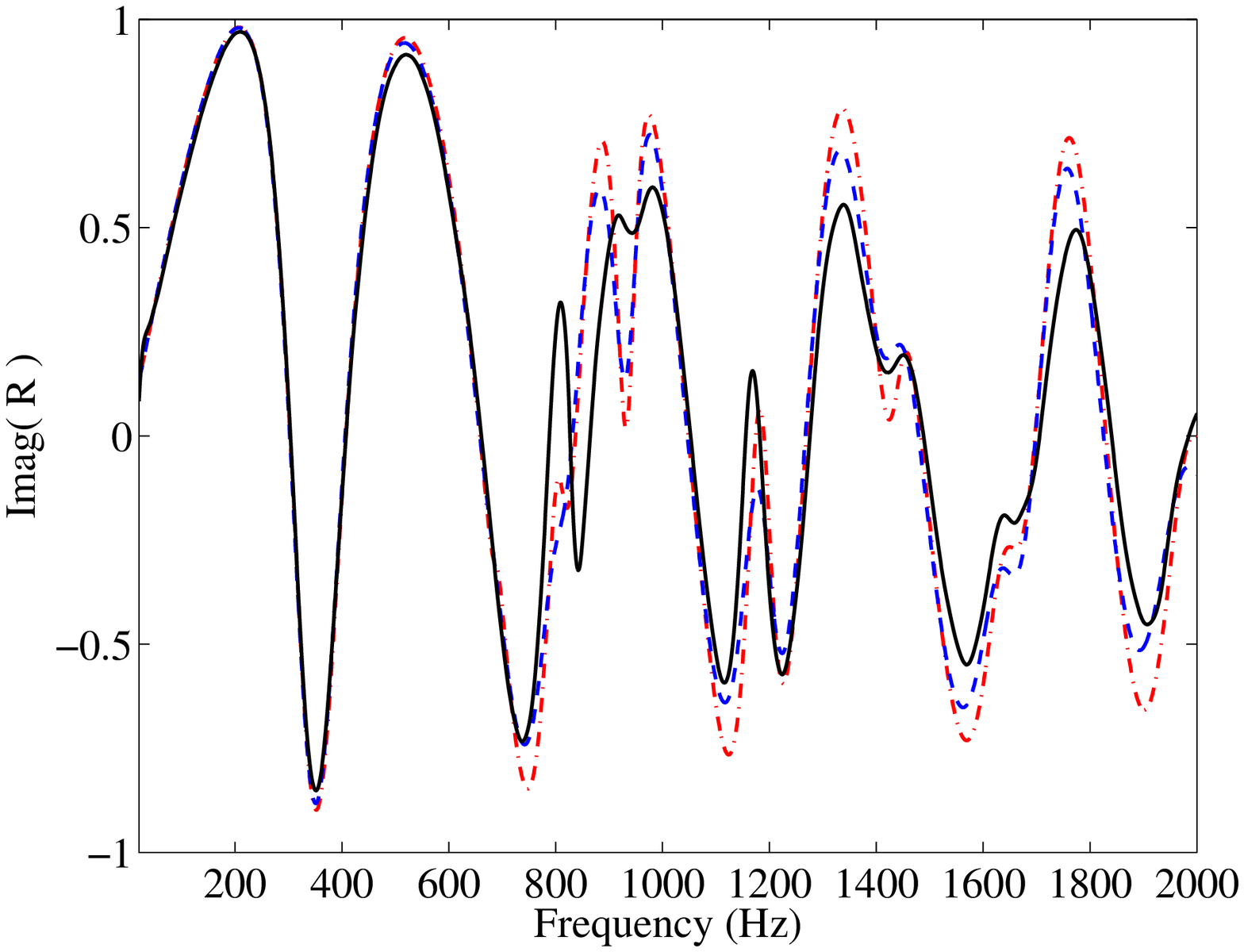} %
\includegraphics[width=246pt]{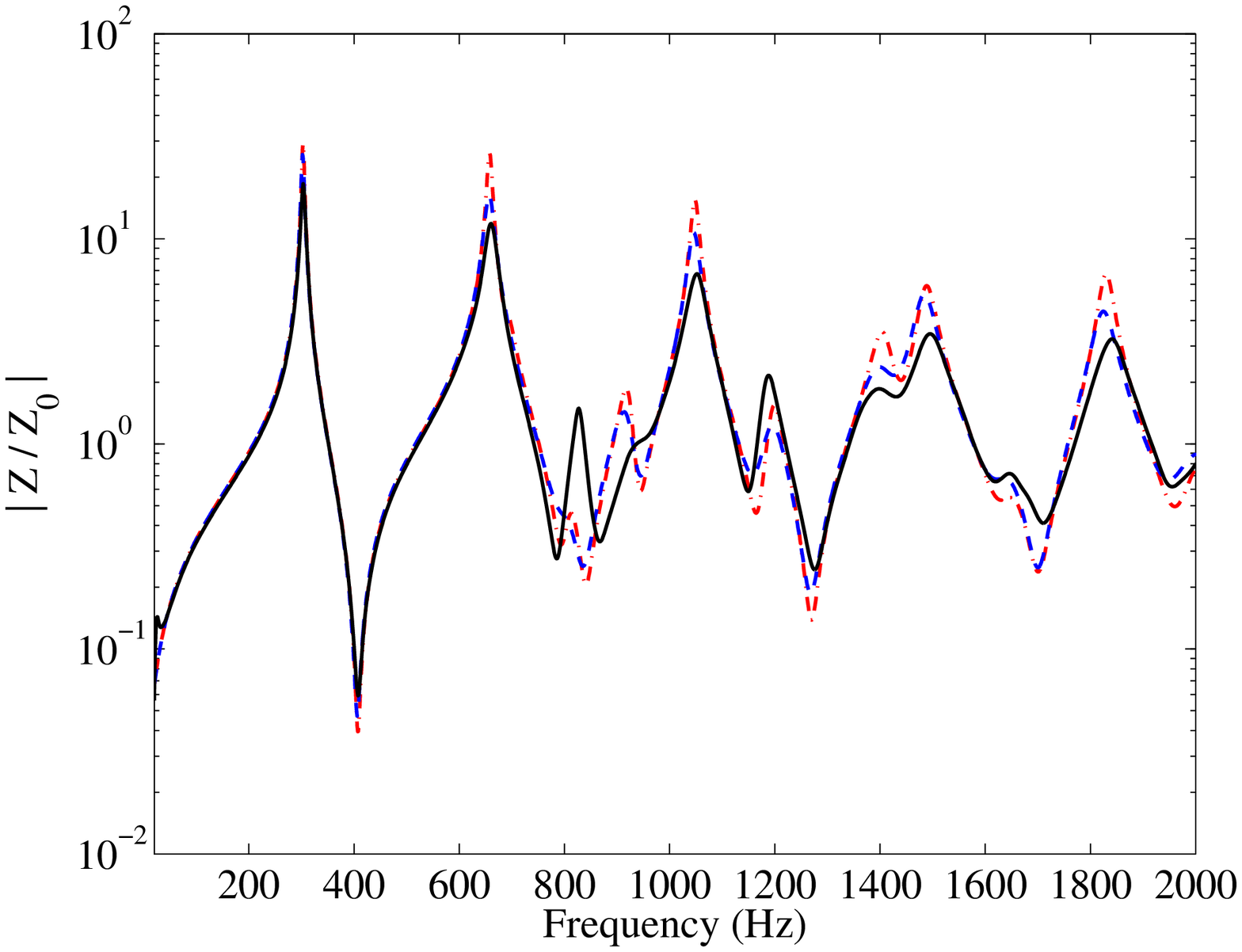}
\end{center}
\caption{Imaginary part of the reflectance (left) and magnitude of the
  impedance (right) of an alto saxophone with a $B\flat$ cross
  fingering: experimental results (black solid) calculations with external
  interactions (blue dashed) and calculations with the TMM (red
  dash-dotted).}
\label{fig:saxo_Bflat}
\end{figure}

\begin{figure}[h]
\begin{center}
\includegraphics[width=246pt]{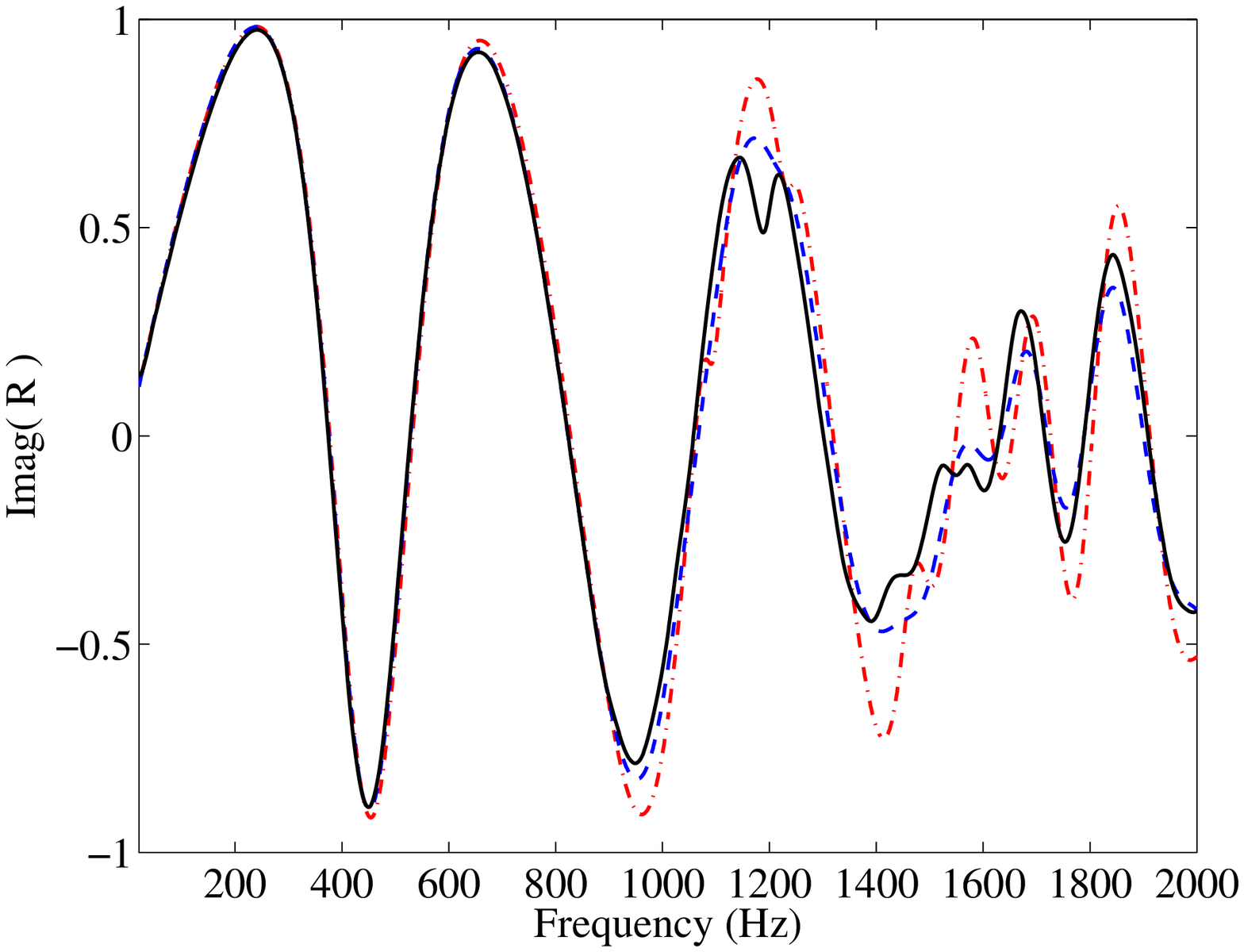} %
\includegraphics[width=246pt]{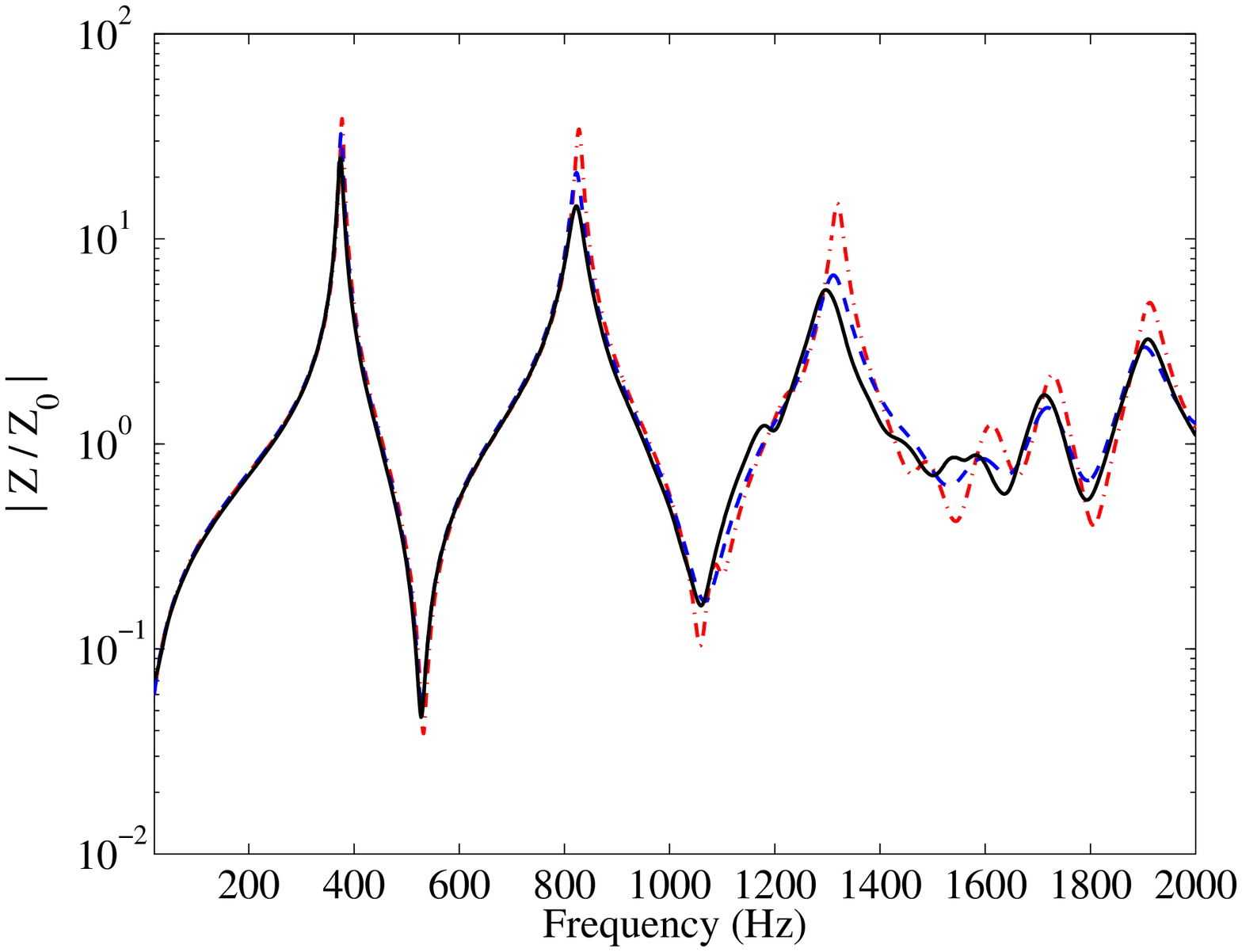}
\end{center}
\caption{Imaginary part of the reflectance (left) and magnitude of the
  impedance (right) of an alto saxophone with a $C\sharp$ fingering:
  experimental results (black solid) calculations with external interactions
  (blue dashed) and calculations with the TMM (red dash-dotted).}
\label{fig:saxo_Csharp}
\end{figure}

The input impedance of an alto saxophone was measured and compared
with calculations using the TMMI and classical TMM methods. The
instrument is a Selmer Super Action Series II, serial \#438024. The
imaginary part of the reflectance and the magnitude of the impedance
for the first register written $B_{3}$, $Bb_{4}$, and $C\#_{5}$
fingerings (respectively $146$, $277$ and $330$\,Hz) are shown in
Figs.~\ref{fig:saxo_B}, \ref{fig:saxo_Bflat} and
\ref{fig:saxo_Csharp}.  These three fingerings correspond to having a
single open tonehole near the bell, a cross-fingering with several
holes closed between open holes, and most holes open,
respectively.  Discrepancies between the experimental data and the
calculations in Fig.~\ref{fig:saxo_B} for frequencies above about 600
Hz are likely due to inaccuracies of the bell geometry and model.

As expected, the TMM and TMMI results are nearly identical when few
holes are open (Fig.~\ref{fig:saxo_B}).  As more holes are opened,
variations are more apparent and the TMMI tends more toward the
experimental results.  The magnitude of the impedance peaks are
generally reduced by the external interactions.  Though difficult to
discern in the figures, the resonance frequencies are slighly lower in
the TMMI results compared to the TMM.

%The behavior of the imaginary part of the reflectance is also better predicted by the TMMI.

\subsection{Clarinet}

\begin{figure}[htb]
\begin{center}
  \includegraphics[width=246pt]{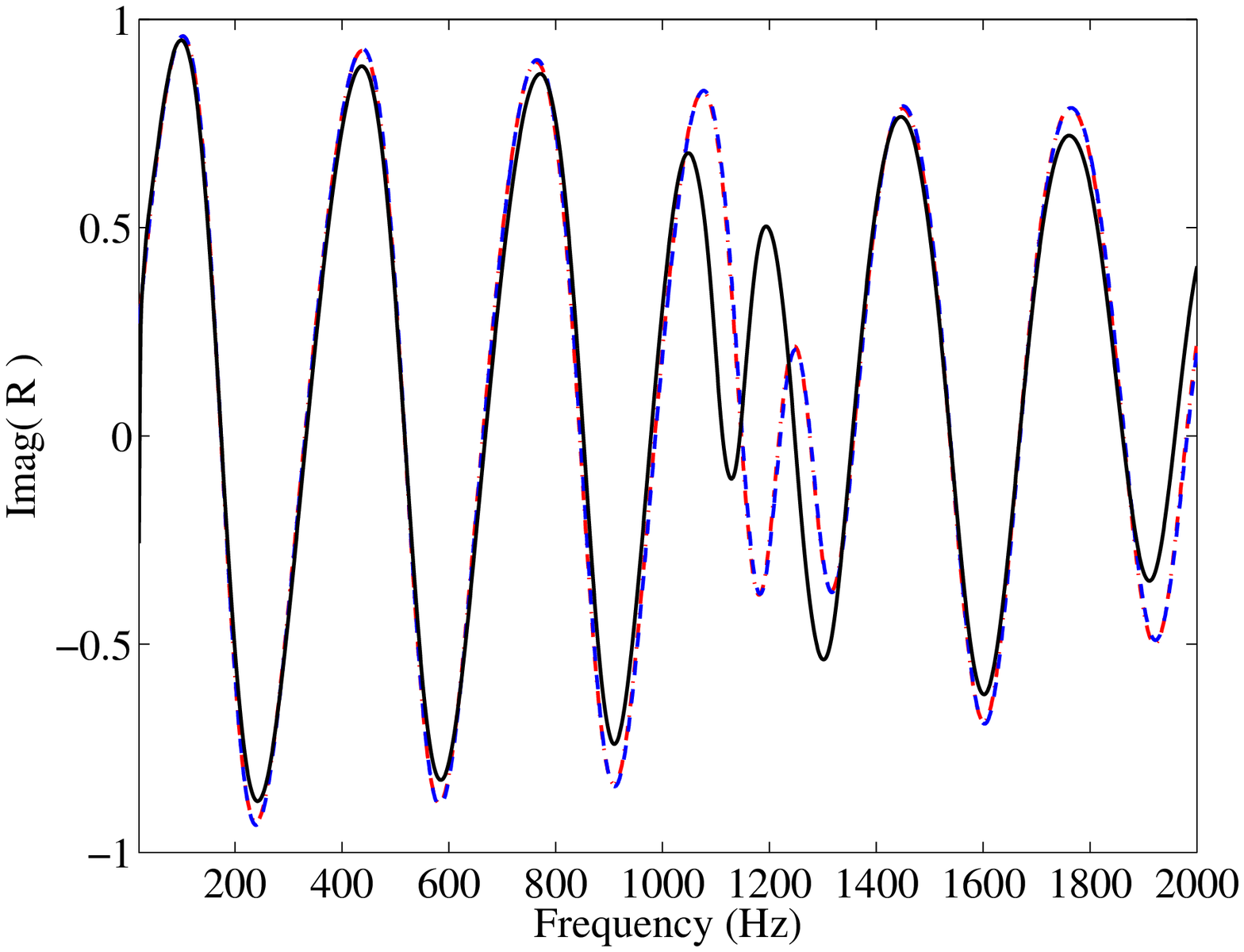} %
  \includegraphics[width=246pt]{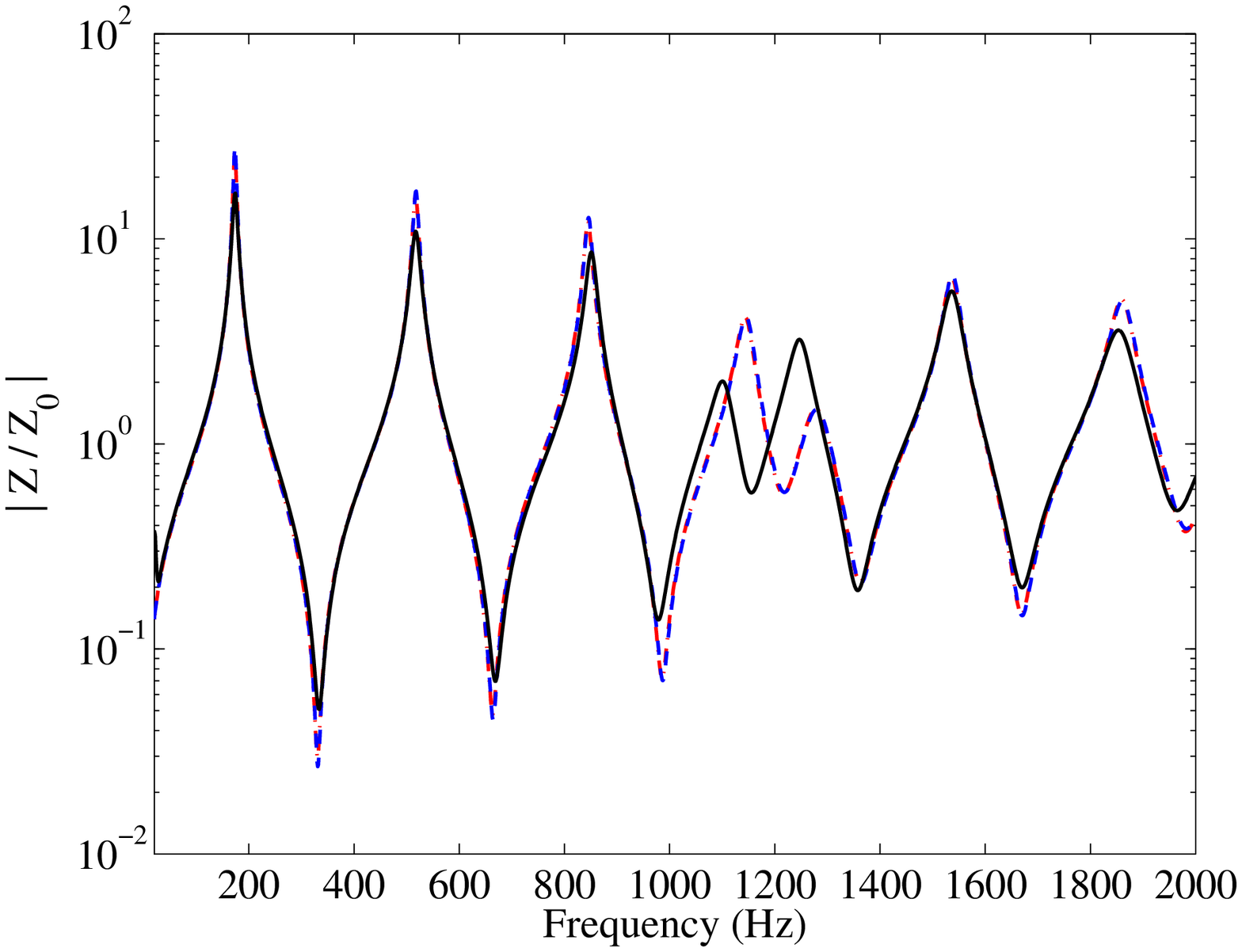}
\end{center}
\caption{Imaginary part of the reflectance (left) and magnitude of the
  impedance (right) of a clarinet with a low $F$ fingering:
  experimental results (black solid) calculations with external
  interactions (blue dashed) and calculations with the TMM (red
  dash-dotted). Because of there are few open holes, the interaction
  effects are very small and the two curves TMM and TMMI are barely
  distinguishable.}
\label{fig:clar_F}
\end{figure}

\begin{figure}[htb]
\begin{center}
\includegraphics[width=246pt]{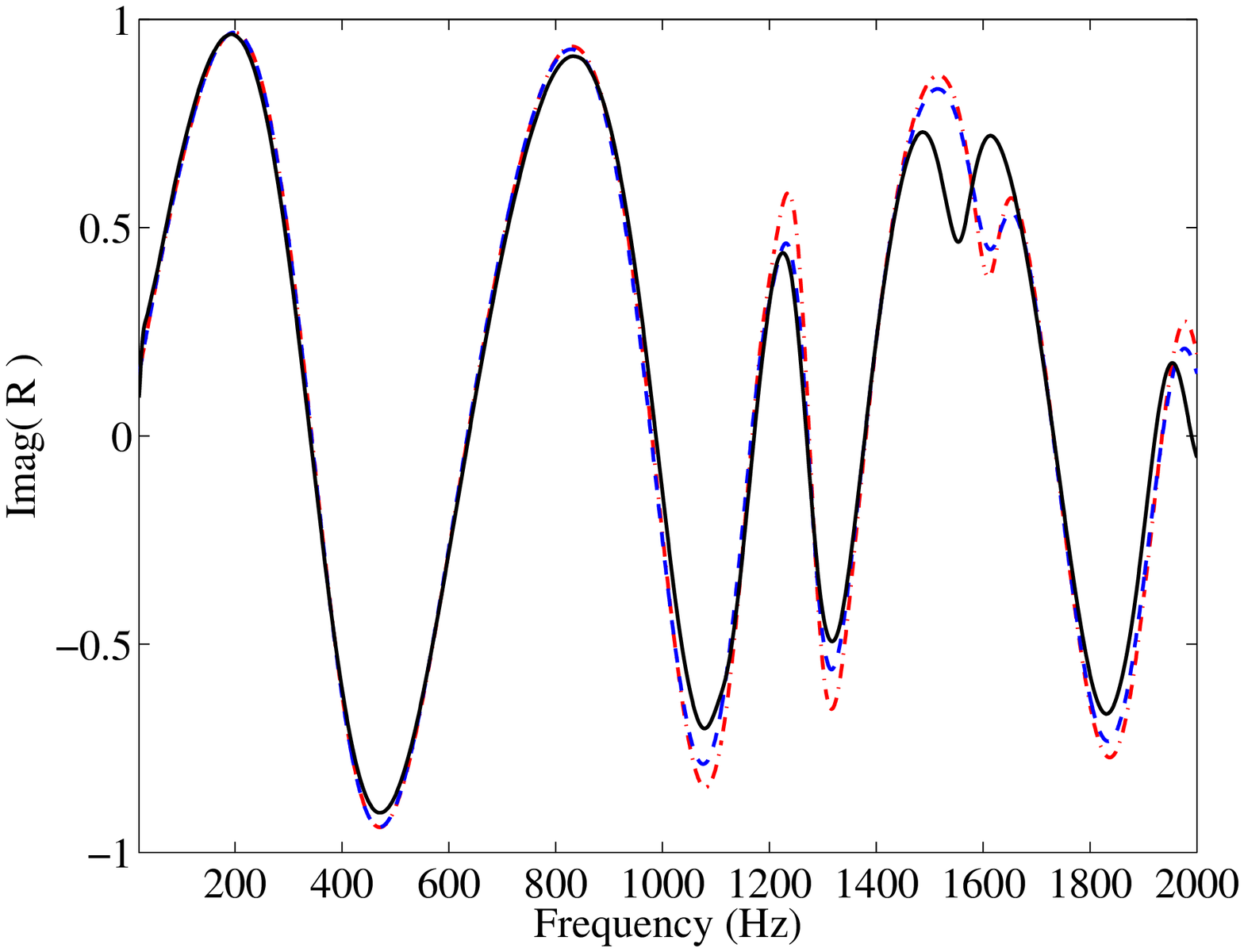} %
\includegraphics[width=246pt]{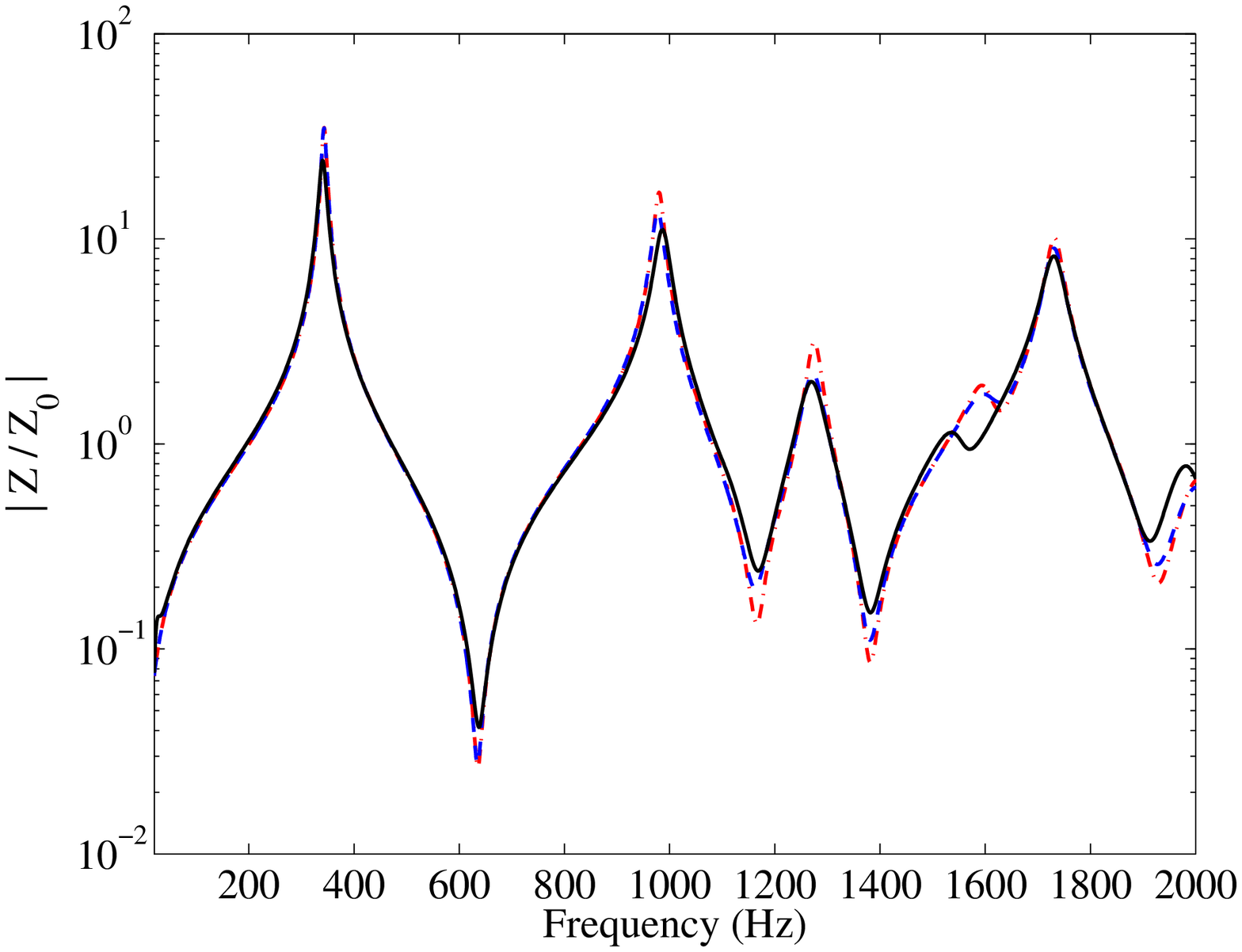}
\end{center}
\caption{Imaginary part of the reflectance (left) and magnitude of the
  impedance (right) of a clarinet with an $E\flat$ cross fingering:
  experimental results (black solid) calculations with external interactions
  (blue dashed) and calculations with the TMM (red dash-dotted).}
\label{fig:clar_Eb}
\end{figure}

\begin{figure}[htb]
\begin{center}
\includegraphics[width=246pt]{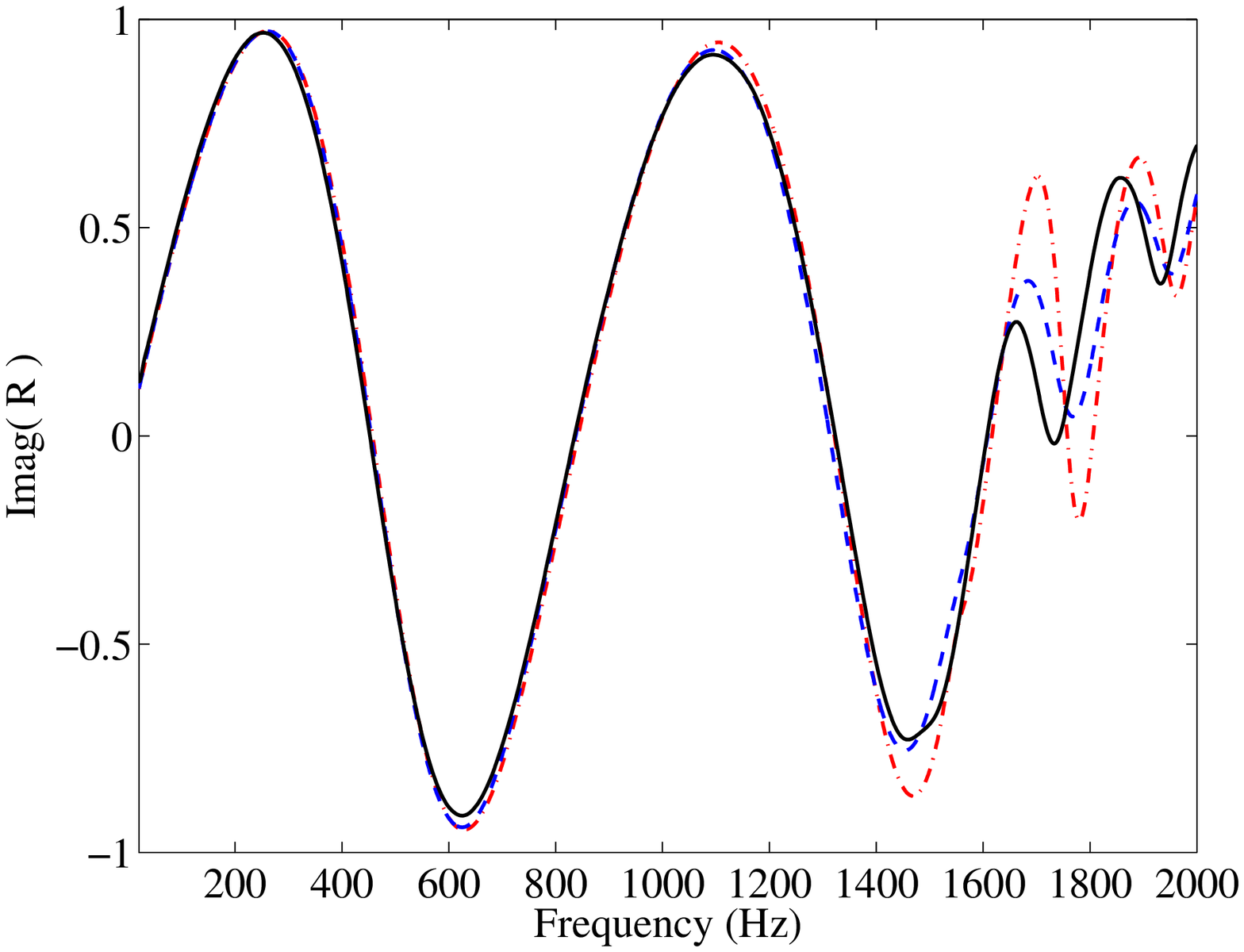} %
\includegraphics[width=246pt]{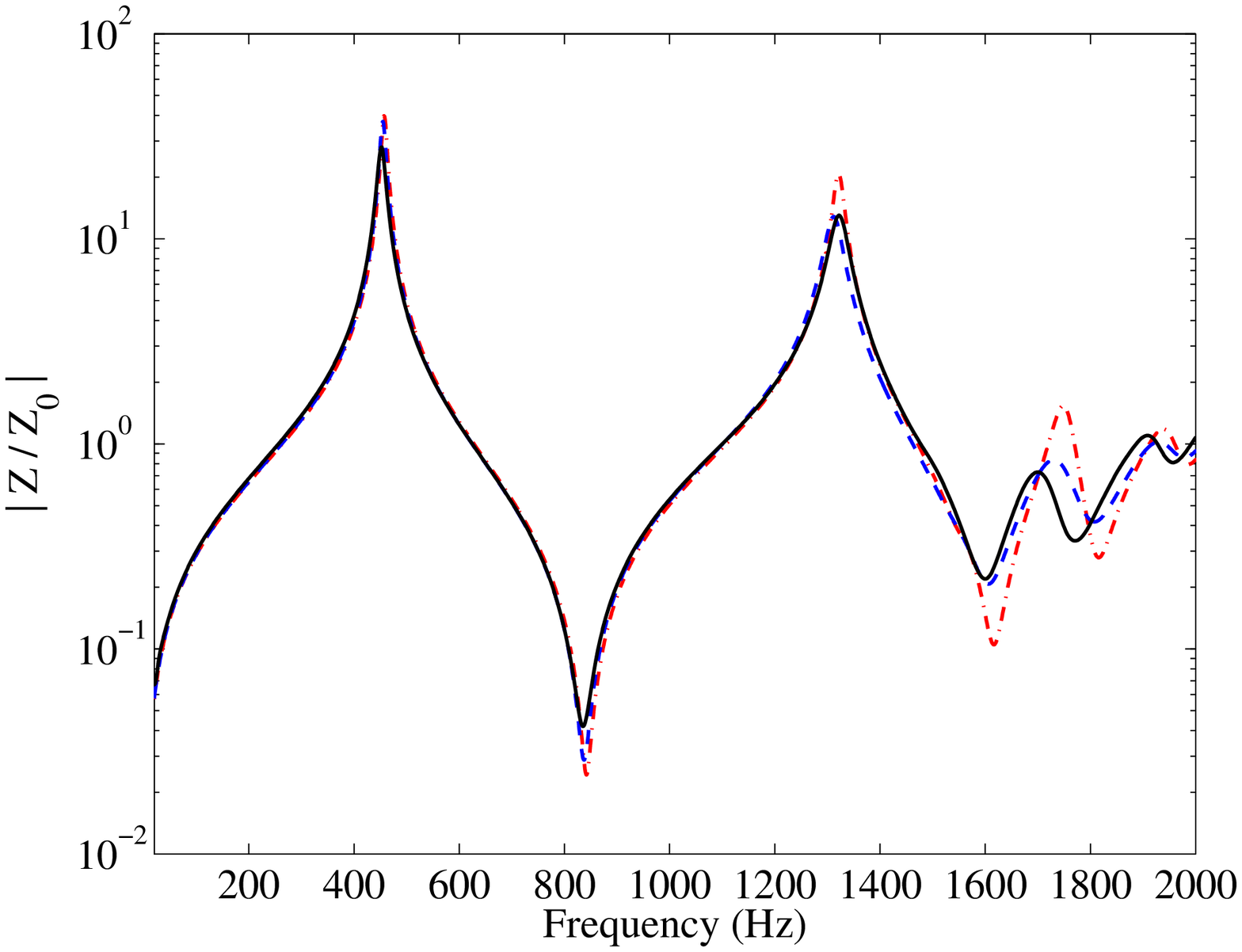}
\end{center}
\caption{Imaginary part of the reflectance (left) and magnitude of the
  impedance (right) of a clarinet with a $G$ fingering: experimental
  results (black solid) calculations with external interactions (blue
  dashed) and calculations with the TMM (red dash-dotted).}
\label{fig:clar_G}
\end{figure}

The input impedance of a $B\flat$ clarinet was measured and compared
with calculations using the TMMI and classical TMM methods. The
instrument is a Selmer USA Signet 100, serial \#211240. The imaginary
part of the reflectance and the magnitude of the impedance for the
first register written $F_{3}$, $E\flat_{4}$, and $G_{4}$ fingerings
(respectively $156$, $277$ and $349$\,Hz) are shown in
Figs.~\ref{fig:clar_F}, \ref{fig:clar_Eb} and \ref{fig:clar_G}.  As
with the saxophone, these three fingerings correspond to having a
single open tonehole near the bell, a cross-fingering with several
holes closed between open holes, and most holes open, respectively.

Similarly to the case of the saxophone, the resonance frequencies of
the clarinet are predicted to be lower when external interactions are
accounted for. For fingerings where many toneholes are open, the
lowering is on the order of 5--10 cents, which is slightly larger than
for the saxophone. As expected, the lowest notes of the instrument,
where only a few toneholes are open, are not much affected.

For higher frequencies, the behavior of the instrument changes more
significantly. As an example, Fig.~\ref{fig:clar_G} displays the
imaginary part of the reflectance and the magnitude of the impedance
for the fingering $G_{4}$ ($349$ Hz) (no fingers down). For the first
two resonances, the external interaction only slightly shifts the
frequencies to a lower value and the maxima of the impedance
corresponds with the zeros of the imaginary part of the
reflectance. Between 1600--2000\thinspace Hz, however, the impedance
magnitude extrema are clearly attenuated by the external interactions
and the resonance frequencies more visibly shifted lower.

\section{Conclusion\label{conclusion}}

The TMMI method provides a more accurate means for the calculation of the
acoustics properties of woodwind instrument than the classical TMM, because
it accounts for external interactions. As explained in Appendix B, it is not
possible to use a simple perturbation of the TMM, but the implementation of
the TMMI is rather easy, and the computation time is short. The
discrepancies between the TMMI and FEM are rather small and can probably be
explained by several factors: first, the values of the radiation-matrix
elements are roughly approximated; second, when several adjacent toneholes
are closed, further improvement of the higher frequency modeling of a
woodwind instrument would require internal coupling of higher-order modes to
be accounted for, at least for the determination of the resonance
frequencies (notice that these effects do not yield radiation effects, thus
dissipative effects, in comparison to the external interactions).

\ Future work needs to be done for a systematic comparison between theory
and experiment for the case of woodwind instruments. This is a long and
delicate task because it requires very precise geometrical measurements,
including bends and positions of the keys over the tone holes, as well as a
precise theory.

Finally, we can summarize some effects of the interactions between holes as:

\begin{itemize}
\item The effect of tonehole interactions is generally more important when
the toneholes are closer together.

\item The order of magnitude of the interaction effect seems to be of the
same order for saxophones and clarinets.

\item At low frequencies, the effect of interaction is of the order of
magnitude of 10 cents when several holes are open, i.e. more than the
threshold interval that the ear can perceive (about 4 cents). This is not negligible for instrument-making purpose.

\item At higher frequencies, the high-pass filtering behavior of the
tonehole lattice allows more flow past the first tonehole and increases the
effect of interactions, particularly near and above the cutoff frequency.

\item Near and above cutoff, the standing wave ratio is reduced by the
effect of interactions. Therefore the effect is important mainly at higher
frequencies, and needs to be taken into account for sound simulation or
synthesis purposes.
\end{itemize}

\section*{Acknowledgements}

This project was funded in part by a Discovery Accelerator Supplements grant from the Natural Sciences and Engineering Research Council of Canada.  The authors also wish to acknowledge the contribution of experimental data by Helmhut Springer during an internship in Le Mans, France in 1986.

%\References{}{fem}{acalit2e}
%\bibliographystyle{apalike}

%\bibliographystyle{plain}
%\bibliographystyle{plain}
%\bibliographystyle{plain}
\bibliography{refs}

\section*{Appendix A: General modeling of an open tonehole}

A general model for an open tonehole can be found in Refs. \cite{dubos_1999,
dalmont_experimental_2002}. It includes (negative) inertances in series,
which can be concatenated with the transfer matrix of the main tube (Eq. (%
\ref{M2})), and a shunt impedance, which can be written as:
\begin{equation}
Z_{s}=jZ_{0h}\left\{ kt_{i}+\tan \left[ k(t+t_{m}+t_{r})\right] \right\},
\label{M5}
\end{equation}%
where $Z_{0h}$ is the characteristic impedance of the tonehole. This includes the effect of the internal added mass, proportional to $t_{i}$%
, the effect of propagation of the planar mode over the length of the hole
chimney $t$, with length corrections corresponding to a matching volume $%
t_{m}$ and to the radiation $t_{r}$, where:%
\begin{equation}
t_{r}=\arctan (Z_{R}/(jZ_{0h}))/k.
\end{equation}%
$Z_{R}$ is the radiation impedance. In order to generalized this model to account for
external interactions, it is necessary to distinguish
the acoustics quantities at the input of a hole (without index) and at its
output (index $rad$), and to write a transfer matrix:%
\begin{equation}
\left(
\begin{array}{c}
p \\
u%
\end{array}%
\right) =\left(
\begin{array}{cc}
c_{t}-kt_{i}s_{t} & jZ_{0h}(s_{t}+kt_{i}c_{t}) \\
Z_{0h}^{-1}js_{t} & c_{t}%
\end{array}%
\right) \left(
\begin{array}{c}
p \\
u%
\end{array}%
\right) ^{rad}
\end{equation}%
where $c_{t}=\cos \left[ k(t+t_{m})\right] ,$ $s_{t}=\sin k(t+t_{m}).$ This
matrix can be written for each hole, and allows the following
matrix relationship to be defined:%
\begin{equation}
\left(
\begin{array}{c}
\mathbf{P} \\
\mathbf{U}%
\end{array}%
\right) =\left(
\begin{array}{cc}
\mathbb{A} & \mathbb{B} \\
\mathbb{C} & \mathbb{D}%
\end{array}%
\right) \left(
\begin{array}{c}
\mathbf{P} \\
\mathbf{U}%
\end{array}%
\right) ^{rad}
\end{equation}%
The equation $\mathbf{U}^{s}\mathbf{=U+}\mathbb{Y}\mathbf{P}$, obtained from
Eqs. (\ref{eq:uleftright}) and (\ref{eq:us}), can be written as follows:%
\begin{equation}
\mathbf{U}^{s}\mathbf{=}(\mathbb{C+YA}\mathbf{)P}^{rad}+(\mathbb{D+YB})%
\mathbf{U}^{rad}  \label{M7}
\end{equation}

If the total length of the tonehole is assumed to be shorter than the
wavelength, $\mathbb{A=D\simeq I}$, $\mathbb{C=}\mathbf{0}$, and $%
B_{nn}=jZ_{0h}k(t+t_{m}+t_{i}).$ Thus, using Eq. (\ref{eq:pu}), Eq. (\ref{M7}%
) leads to:%
\begin{equation}
\mathbf{U}^{s}\mathbf{=}(\left[ \mathbb{I+Y(Z+B)}\right] \mathbf{U}
\label{M8}
\end{equation}

%\section*{\protect\bigskip \protect\bigskip Appendix B: Is it Possible to
%Compute the External Interaction by the Transfer Matrix Method?}

\section*{Appendix B: Is it Possible to Compute the External Interaction by the Transfer Matrix Method?}

We consider the equation to be solved:%
\begin{equation}
\mathbf{U=(1+YZ}_{R}\mathbf{)}^{-1}\mathbf{U}_{s}  \label{KK}
\end{equation}
It is interesting to study if it is possible to solve this equation by
perturbation, starting from the TMM method. A quite natural way to do this
is to consider that the effect of the external interaction is weak, and to
keep a calculation based upon transfer matrices. A first calculation is done
without interaction, then the pressures are modified by calculating them
with interactions taken into account. The perturbation calculation can be
stopped here, but it is possible to iterate it: a new self-impedance is
calculated as the ratio of the modified pressure to the unmodified flow
rate, then the new flow rates can be calculated again from the transfer
matrix method with the modified values of the self-impedances. In practice,
the iteration scheme is found to converge for almost all frequencies except
low ones. This result is intuitive because in the stop band, the external
sound pressure decreases proportionally to the inverse of the distance,
while the internal pressure decreases exponentially, therefore the external
interaction is more significant.

It is possible to derive a criterion of convergence for the iteration
procedure and, when it converges, it is possible to prove that the result is
correct. This is done hereafter. At each step $n$ of the calculation, the
transfer matrix method leads to the following relationship between the
source $\mathbf{U}_{s}$, having a single non-zero element, $U_{s}(1)$, and
the pressure and flow rate vectors, $\mathbf{P}^{(n)}$ and $\mathbf{U}^{(n)}$%
:%
\begin{equation}
\mathbf{U}_{s}\mathbf{=U}^{(n)}\mathbf{+YP}^{(n)}\text{.}  \label{KK2}
\end{equation}%
The calculation is done by defining a diagonal matrix for the termination
impedance of each hole (both the direct method and the transfer matrix
method can be used):
\begin{equation}
\mathbf{P}^{(n)}\mathbf{=D}^{(n)}\mathbf{U}^{(n)}.  \label{KK3}
\end{equation}%
From the knowledge of the flow rate $\mathbf{U}^{(n)}$, the next value of
the pressure $\mathbf{P}^{(n+1)\text{ }}$is deduced:%
\begin{equation}
\mathbf{P}^{(n+1)}\mathbf{=Z}_{R}\mathbf{U}^{(n)}.  \label{KK4}
\end{equation}%
The iteration equation is therefore found to be, with $\mathbf{M=YZ}_{R}$:%
\begin{equation}
\mathbf{U}^{(n+1)}\mathbf{=U}_{s}\mathbf{-MU}^{(n)}.  \label{KK5}
\end{equation}%
The recurrence relationship leads to the following solution:%
\begin{equation}
\mathbf{U}^{(n)}=\left[ \overset{n-1}{\underset{i=0}{\sum }}(-1)^{i}\mathbf{M%
}^{i}\right] \mathbf{U}_{s}+(-1)^{n}\mathbf{M}^{n}\mathbf{U}^{(0)}.
\label{KK6}
\end{equation}%
If the norm of the matrix $\mathbf{M}$\textbf{\ }is less than unity, the
recurrence converges to the solution (\ref{KK}), the series corresponding to
a Neumann series expansion.

Now the starting point can be discussed. The first idea is to deduce the
solution without interaction from the transfer matrix product :%
\begin{equation}
\mathbf{U}^{(0)}\mathbf{=(1+YD)}^{-1}\mathbf{U}_{s}\text{,}  \label{KK7}
\end{equation}%
where $\mathbf{D}$ is the diagonal matrix of the self-impedances of $\mathbf{%
Z}_{R}.$ Another possibility is to start with $\mathbf{U}^{(0)}\mathbf{=U}%
_{s}$: this implies that the first pressure vector is built with the
pressures created by a flow rate located at the first open hole. It can be
concluded that the transfer matrix method can be used when the norm of the
matrix $\mathbf{YZ}_{R}$ is less than unity. Because it can be verified that
this is not true in the stop band, the perturbation method unfortunately
cannot be used in general. This confirms the intuition: looking at Fig. \ref%
{fig:proto1_R}, it appears that the effect of external interaction can be
very large in stop bands for holes very far apart from each other and the
perturbation method cannot converge.

Nevertheless, in pass bands we observe that convergence occurs rapidly when
starting from Eq. (\ref{KK7}), and even the first order, corresponding to a
single perturbation step, is satisfactory. This observation thus justifies
the reasoning given in Sec.~\ref{results}.

\end{document}